\documentclass[12pt,epsf,amssymb]{article}
\usepackage{tabularx}
\usepackage{array}
\usepackage{graphics}
\usepackage{graphicx}
\usepackage{epsfig}
\usepackage{amsmath}
\usepackage{amssymb}
\usepackage{citesort}
\makeatletter

\usepackage{verbatim}

\newcommand{\msbar}{{\overline{\rm MS}}}

\newcommand{\bea}{\begin{eqnarray}}
\newcommand{\eea}{\end{eqnarray}}
\newcommand{\beq}{\begin{equation}}
\newcommand{\eeq}{\end{equation}}
\newcommand{\gev}{{\rm GeV}}
\newcommand{\err}{{\rm err}}
\newcommand{\mev}{{\rm MeV}}

\newcommand{\pdir}{p\kern -5.2pt\raise 0.2ex\hbox {/}}
\newcommand{\vdir}{v\kern -5.75pt\raise 0.15ex\hbox {/}}
\newcommand{\kdir}{k\kern -5.75pt\raise 0.15ex\hbox {/}}
\newcommand{\epsdir}{\epsilon\kern -5.0pt\raise 0.15ex\hbox {/}}
\newcommand{\bvdir}{\bar{v}\kern -5.75pt\raise 0.15ex\hbox {/}}
\newcommand{\Ddir}{D\kern -7.75pt\raise 0.20ex\hbox {/}}
\newcommand{\ldir}{l\kern -5.0pt\raise 0.2ex\hbox{/}}
\newcommand{\varepsdir}{\varepsilon\kern -5.5pt\raise 0.15ex\hbox{/}}

\def\str{\operatorname{str}}
\def\tr{\operatorname{tr}}
\def\Tr{\operatorname{Tr}}

\def\diag{\operatorname{diag}}
\def\negcdot{\negmedspace\cdot\negmedspace}
\def\bpi{B \to \pi \ell \nu}

\makeatother

\begin{document}
\thispagestyle{empty} 
\begin{flushright}
\begin{tabular}{l}
{\tt BNL-HET-02/16}\\
{\tt IJS/TP-28/02}\\
{\tt Rome-1340/02}\\
\end{tabular}
\end{flushright}
\begin{center}
\vskip 2.2cm\par
{\par\centering \LARGE \bf  {\boldmath $B\to \pi$ and $B\to K$} transitions}\\   
{\phantom{\huge{}}}
{\par\centering \LARGE \bf in standard and quenched}  \\ 
{\phantom{\huge{}}}
{\par\centering \LARGE \bf chiral perturbation theory} 
\vskip 1.15cm\par
{\par\centering \large  
\sc Damir~Be\'cirevi\'c$^a$, Sa\v{s}a Prelov\v{s}ek$^{b,c,d}$ and Jure~Zupan$^c$}
{\par\centering \vskip 0.5 cm\par}
{\sl 
$^a$Dip. di Fisica, Universit\`a di Roma ``La Sapienza",\\ 
Piazzale Aldo Moro 2, I-00185 Rome, Italy. }\\
{\par\centering \vskip 0.3 cm\par}
{\sl 
$^b$Theory Group, Brookhaven National Laboratory,\\ 
Upton, NY 11973, USA. }\\
{\par\centering \vskip 0.3 cm\par}
{\sl 
$^c$J.~Stefan Institute, Jamova 39, P.O. Box 3000,\\ 
1001 Ljubljana, Slovenia.}\\
{\par\centering \vskip 0.3 cm\par}
{\sl 
$^d$Departement of Physics, University of Ljubljana, Jadranska 19,\\ 
1000 Ljubljana, Slovenia.}\\

{\vskip 0.75cm\par}
\end{center}

\begin{abstract}
We study the effects of chiral logs on the $\text{\it heavy} \to \text{\it light}$ 
pseudoscalar meson transition form factors by using standard 
and quenched chiral perturbation theory combined with the static 
heavy quark limit. The resulting expressions are used to indicate  
the size of uncertainties due to the use of the quenched approximation 
in the current lattice studies. They may also be used to assess 
the size of systematic uncertainties induced by missing chiral 
log terms in extrapolating toward the physical pion mass. 
We also provide the  coefficient multiplying the quenched chiral log, which
 may be useful if the quenched lattice studies 
are performed with very light mesons. 
\end{abstract}
                 \vskip 2.2 cm 
\setcounter{page}{1}
\setcounter{footnote}{0}
\setcounter{equation}{0}
\noindent

\renewcommand{\thefootnote}{\arabic{footnote}}

\newpage
\setcounter{footnote}{0}

\section{Introduction}
\setcounter{equation}{0}
Over the past decade, a considerable amount of effort has been put into studying the
nonperturbative QCD dynamics of the $\text{\it heavy}\to \text{\it light}$ decays. The main 
target was (and still is) the extraction of the CKM matrix element 
$\vert V_{ub}\vert$. The prerequisite for its determination from 
the exclusive $ \bpi$ decay mode is a precise knowledge of the 
relevant form factors. Accurate information on the form factors 
is crucial also when  studying  the impact  
of physics beyond the standard model (SM) on the exclusive $b\to s \ell^+ 
\ell^-$  modes.
 
The fact that the kinematically accessible region of momentum transfers is very large 
(e.g., for $\bpi$ it is $0\leq q^2 \leq 26.4~\gev^2$) makes  QCD-based calculations 
of form factors ever more difficult. 
The physical pictures emerging at the two extremities of the $q^2$ region are quite 
different and effective field theory approaches, based on the appropriate use of
the heavy quark expansion,  have been developed to simplify the description 
of these processes. The heavy quark effective theory (HQET), which is applicable for 
recoil momenta close to zero ($q^2 \to q^2_{\rm max}$),~\footnote{Zero recoil refers to 
the recoil of the daughter meson in the rest frame of the decaying one. 
In $\bpi$, it means that the pion is soft.} provides us with valuable scaling laws 
for the form factors~\cite{isgur}. In the region of large recoils ($q^2 \to 0$), instead, 
the large energy effective theory and its descendants help resolve the heavy quark 
dependence of the form factors~\cite{orsay,bauer}. Although these conceptual 
steps forward are highly beneficial for a better understanding of the underlying dynamics, 
a model-independent description (calculation) of the form factors in the entire physical 
region is still missing.

Among the QCD-based approaches employed to compute the $\text{\it heavy}\to \text{\it light}$ 
decay form factors, the following two stand apart.
\begin{itemize}
\item[$\otimes$] Light cone QCD sum rules (LCSR). This analytic approach 
 contains the least number of assumptions and has the correct heavy quark mass scaling properties.   
The range of applicability is, however, limited to low $q^2$'s~\cite{khodjamirian}.  
\item[$\otimes$] Lattice QCD. This method allows us to solve the nonperturbative QCD effects numerically. 
Because of the current insufficient computing power, the $B\to \pi$ transition is reached either  
(a) by extrapolating the directly computed form factors from the region around 
the charm to the $b$-quark mass~\cite{ukqcd,ape}, or 
(b) by using a latticized effective 
theory of the heavy quark, such as NRQCD~\cite{jlqcd} (see also ref.~\cite{shigemitsu}), or 
(c) by reinterpreting lattice QCD in terms of NRQCD when the heavy quark mass becomes larger than 
the lattice UV cutoff~\cite{fnal}.
All these strategies share one feature: the accessible form factors are restricted 
to the region of small recoils.
\end{itemize}
It is fair to say that the LCSR and lattice QCD are complementary 
to each other; it is important to use them both in order to check their consistency 
and from their comparison perhaps learn more about the underlying 
nonperturbative QCD dynamics.

Since the lattice studies are expected to provide us with the most 
accurate predictions about the shapes and absolute values of the form 
factors, it is important to have good control over the various
assumptions that are currently used in lattice simulation and the data analysis. 
Two sources of systematic uncertainty have so far been ignored: 
the quenched approximation and possible deviations from the 
linear or quadratic chiral extrapolation forms.

All the available lattice results for $\bpi$ decay form factors 
are obtained from simulations in the quenched approximation, where
 the sea quark loop effects in the QCD vacuum fluctuations are neglected ($n_F=0$). 
To get an idea about the systematic error induced by the quenched approximation, one can confront the 
expressions for the form factors derived in the standard and in  quenched 
chiral perturbation theory (ChPT and QChPT, respectively). Such expressions, 
at leading order in the heavy quark expansion  and  
 next-to-leading order (NLO) in the chiral expansion, are provided in
the present paper.  These expressions are also useful in 
assessing the systematic uncertainties due to chiral extrapolations. 
Current lattice studies deal with light mesons of masses $\gtrsim 450$~MeV. 
The physical pion mass is reached through a linear or quadratic 
extrapolation in the light quark mass. Although it is not clear for which 
light quark masses one begins to probe the subtleties of the chiral expansion, 
it is beyond reasonable doubt that, very 
close to the chiral limit, chiral log terms of the form $m_\pi^2 \log(m_\pi^2)$ 
may modify the result of the extrapolation. The coefficients multiplying the chiral 
logs are predicted by (Q)ChPT and will be presented in this paper. 
Finally, in the case of $B\to K$ decay the standard lattice 
strategy is to consider the kaon as a meson consisting of degenerate quark masses. 
The impact of nondegeneracy in the quenched approximation may also be addressed 
by using the QChPT expressions for the form factors, as we shall see later on.

It is important to stress a difficulty in getting reliable numerical 
estimates from this approach. As we just mentioned, it is not clear 
for what value of the light meson mass the chiral logs become relevant. That ambiguity is 
important since one extrapolates from the heavier light masses, for which  
chiral logs have {\it not} been observed. Another obstacle is the multitude of 
low energy constants that appear in the Lagrangian and in the transition operators 
in both quenched and unquenched ChPT. The values of some of those constants 
are unknown or simply guessed. Furthermore, as we shall see, the appearance of 
the chirally divergent  quenched chiral logs obliges us to compare the full and 
quenched expressions for not-so-light mesons, for which the ChPT is less predictive. 
For these three reasons, the numerical results inferred from this approach should 
always be taken with a grain of salt. In other words, rather than true estimates  
of the quenching errors, our numerical results should be considered as a mere indicator 
of the size of those errors.

The rest of the paper is organized as follows. In Sec.~\ref{Definitions} 
we recall the basic elements of (Q)ChPT and its combination with the leading order 
HQET Lagrangian; in addition to standard definitions of the form factors, we will 
introduce some that are more convenient for our purposes; in 
Sec.~\ref{sec:chi-results} we present the one-loop (standard and quenched) 
chiral corrections for our form factors; in Sec.~\ref{sec:parameters} we 
discuss the values of the low energy constants that we chose for the numerical 
analysis, which we present in secs.~\ref{sec:numerics1} 
and~\ref{sec:numerics2}; we conclude in Sec.~\ref{sec:conclusions}.

\section{Setting the scene\label{Definitions}}

In this section we recall some basic features of QChPT, as it was developed in the 
papers by  Sharpe~\cite{sharpe} and by Bernard and Golterman~\cite{bernard}.~\footnote{
For an elegant alternative way to introduce partially quenched ChPT,  see ref.~\cite{damgaard}.}
Although QChPT resembles standard ChPT in many aspects there are 
important differences. The main one is the presence of the light $\eta^\prime$ state, 
which in  quenched QCD (QQCD) does not decouple from the octet of pseudoscalar mesons. 
As a result the ``pion" propagator exhibits not only a pole structure but 
also a double-pole one, which is the source of the pathology of the quenched 
approximation.

\subsection{Quenched chiral Lagrangian}

In QQCD, in addition to the quarks $q_a$ ($a=1,2,3$), one also has the bosonic ``ghost"  
quarks $\tilde{q}_a$, of spin $\tfrac{1}{2}$ and with identical mass  $m_{\tilde{q_a}}/m_{q_a}=1$.
Their r\^ole is to  cancel the contributions of the closed quark loops, 
i.e., they provide quenching. If one assumes that the symmetry 
breaking pattern of QQCD is similar to that of the full QCD, i.e. 
the graded $SU(3|3)_L\otimes SU(3|3)_R$ spontaneously breaks down 
to $SU(3|3)_V$, the following chiral Lagrangian can be written 
\bea \label{light}
{\cal L}_{\rm light}={f^2\over 8}\str \left(\partial_\mu \Sigma \partial^\mu
\Sigma^\dagger \right) + {f^2 \mu_0\over 2}
\str({\cal M}\Sigma+{\cal M}\Sigma^\dagger)
+\alpha_0\partial_\mu\Phi_0\partial^\mu \Phi_0-
m_0^2\Phi_0^2+ {\cal L}_{4}\;,
\eea
where we adopt the convention that $f \approx 130$~MeV, and the notation 
\bea \label{eqQ:2}
\Sigma = \exp\left(2i\frac{\Phi}{f}\right), 
\qquad \Phi=
\begin{pmatrix}
\phi&\chi^\dagger\\
\chi&\tilde \phi\;
\end{pmatrix}\;.
\eea
The following comments are in order.
\begin{itemize}
\item[--] In addition to the standard ($q\bar{q}$) Goldstone bosons ($\pi, K,\eta$) 
and the $\eta'$ meson,~\footnote{We will neglect the mixing of $\eta$ and $\eta'$ states, 
as it is irrelevant for the discussion that follows.}, 
organized in the $3\times 3$ matrix
\bea
\phi =\left(
\begin{array}{ccc} 
{1\over \sqrt{2} } \pi^0  +  {1\over \sqrt{6} } \eta +{1\over \sqrt{3} }\eta' &  \pi^+  & K^+ \\
 \pi^-    &   -{1\over \sqrt{2} } \pi^0  +  {1\over \sqrt{6} } \eta +{1\over \sqrt{3} }\eta' & K^0\\
K^-& \bar K^0  &-{2\over \sqrt{6} } \eta +{1\over \sqrt{3} }\eta' \\
\end{array}
\right)\;,
\eea
the ghost--ghost ($\tilde q \bar{\tilde{q}}$) bosons ($\tilde \phi$), as well as the pseudoscalar 
fermions $\bar{\tilde{q}} q$ ($\chi^\dagger$) and  $\bar q \tilde q$ ($\chi$), also appear in eq.~(\ref{eqQ:2}).
\item[--] The global symmetry $SU(3|3)_L\otimes SU(3|3)_R$ is graded and in eq.~(\ref{light}), 
instead of the familiar trace, one deals with the supertrace $\str(\Phi)=\tr(\phi)-\tr(\tilde\phi)$.
\item[--] As already stressed, $\eta^\prime$ does not decouple from the light 
pseudoscalar octet. Its effect is included in two terms of
the Lagrangian~(\ref{light}), each of them multiplied by a new low energy constant, namely,  
$\alpha_0$ and $m_0$. Note that $\Phi_0= \str(\Phi)/\sqrt{6}$ $= 
(\eta^\prime-\tilde{\eta^\prime})/\sqrt{2}$.
\item[--] The quark-ghost mass matrix is diagonal  
${\cal M}=\diag(m_u,m_d,m_s,m_u,m_d,m_s)$. After diagonalizing the mass
term in eq.~(\ref{light}), one gets  
\bea \label{gell-mann}
m_\pi^2 = 4 \mu_0 m_q \;,
\quad 
m_K^2 = 2 \mu_0 \left( m_q + m_s \right)\;,
\quad m_\eta^2 = {4\mu_0\over 3} \left( m_q + 2 m_s \right)\;,
\eea
verifying the familiar Gell-Mann--Okubo relation $4 m_K^2 - m_\pi^2 - 3 m_\eta^2 = 0$.
Notice that we neglect the isospin symmetry breaking, i.e., we set $m_u = m_d 
\equiv m_q$.
\item[--] In eq.~(\ref{light}) ${\cal L}_4$  stands for the terms of ${\cal
O}(p^4)$, of which we write only those that are relevant to the  
heavy-to-light form factors, namely,
\beq \label{eqQ:1}
\begin{split}
{\cal L}_4 = 4  \mu_0 \big\{& L_4 \str\left(\partial_\mu \Sigma \partial^\mu 
\Sigma^\dagger\right)\str\left({\cal M}\Sigma^\dagger+\Sigma 
{\cal M}^\dagger\right) \\
&+ L_5 \str\left[\partial_\mu \Sigma^\dagger \partial^\mu 
\Sigma\left({\cal M}\Sigma^\dagger+\Sigma {\cal M}^\dagger\right)\right]+\cdots \big\}\;.
\end{split}
\eeq
$L_4$ and $L_5$ generate the mass correction to the 
decay constants (and to the wave function renormalization constant)~\cite{gasser}.
\end{itemize}
It is straightforward to verify that after setting  
$\str\to \tr$, $\Phi\to\phi$, $\eta'\to 0$, eq.~(\ref{light}) leads to  
the standard (full QCD) chiral Lagrangian.~\footnote{ For a review  
of the standard ChPT see one of the references listed in~\cite{chiral-reviews}.}

\subsection{Incorporating the heavy quark symmetry}

QChPT has been combined with the leading order HQET  in the work 
by Booth~\cite{booth} and by Sharpe and Zhang~\cite{zhang}. They applied the approach to 
compute the  heavy--light decay constants, the $B^0$--$\bar B^0$ mixing parameter and the Isgur--Wise
function~\footnote{For a recent result on the Isgur--Wise function in 
partially QChPT, see ref.~\cite{savage}.}.

To devise a Lagrangian for the heavy--light mesons, it is necessary to include 
the heavy quark spin symmetry. This is achieved by combining the pseudoscalar 
($P^a$) and vector ($P^{\ast\ a}_\mu$) 
heavy--light mesons in one field: 
\bea
H_a(v) = {1 +  \vdir \over 2} \left[ P^{\ast\ a}_\mu (v)\gamma_\mu - P^a (v)\gamma_5
\right]\;,
\eea 
where $(1+ \vdir )/2$ projects out the particle component of the heavy quark only. 
The conjugate field is defined as $\overline H_a(v) = \gamma_0 H_a^\dagger (v) \gamma_0$,  
whereas the covariant derivative and the axial field have the following forms: 
\bea
&&D^\mu_{ba}H_b = \partial^\mu H_a - H_b{\bf V}^\mu_{ba}
=  \partial^\mu H_a -  H_b {1\over 2}[ \xi^\dagger \partial_\mu \xi + 
\xi \partial_\mu \xi^\dagger ]_{ba}\;,\nonumber \\
&&{\bf A}_\mu^{ab}
= {i\over 2}[ \xi^\dagger \partial_\mu \xi - 
\xi \partial_\mu \xi^\dagger ]_{ab} \;.\eea
In the above equation, $a$ and $b$ run over the light quark flavors and $\xi = \sqrt{\Sigma}$. 
With these ingredients in hand, we now write the quenched chiral Lagrangian for the heavy-light 
mesons as 
\bea \label{heavy}
&&{\cal L}_{\rm heavy}=-\str_a \Tr[\overline{H}_a i v \negcdot D_{ba}H_b]+g 
\str_a\Tr[\overline{H}_aH_b \gamma_\mu 
{\bf A}_{ba}^\mu \gamma_5]\cr
&&\hfill \cr
&&\hspace*{1.8cm}+ g'\str_a \Tr[\overline{H}_aH_a \gamma_\mu \gamma_5] 
\str({\bf A}^{\mu})+{\cal L}_3 \;,
\eea
where $g$ ($g^\prime$) is the coupling of the heavy meson doublet to 
the Goldstone boson (to  $\eta^\prime$  or $\tilde \eta^\prime$). 
A term with $g^\prime$  is thus an artifact of the quenched theory. 
The higher order terms in the expansion in $v\negcdot p\sim{\cal O}( p)$, 
${\cal O}(p^2)$, and in $m_q\sim {\cal O}(p^2)$, denoted as ${\cal L}_3$ 
in eq.~(\ref{heavy}), have the following form~\cite{booth}
\bea \label{eqQ:3}
{\cal L}_3 &=&2\lambda_1\str_a \Tr[\overline H_aH_b]
({\cal M_+})_{ba}+k_1\str_a \Tr[\overline{H}_a i v \negcdot D_{bc}H_b]
({\cal M_+})_{ca}\cr
&&\hfill \cr
&&+k_2\str_a \Tr[\overline{H}_a i v \negcdot D_{ba}H_b]\str_c({\cal M_+})_{cc}+\dots\;,
\eea
with ${\cal M_+}=(1/2)(\xi^\dagger M\xi^\dagger+\xi M\xi)$. We displayed only the terms 
that contribute to the heavy-to-light form factors.
In the above equations, ``Tr" stands for the trace over Dirac indices, 
whereas ``str" is the supertrace over the light flavor indices.

As in the previous subsection, one can easily verify that after replacing 
$\str\to \tr$, $\Phi\to\phi$, $\eta'\to 0$, and $g'\to 0$, one recovers 
the standard chiral Lagrangian for heavy-light mesons (for recent reviews see 
ref.~\cite{reviews-HQChPT}, and for the original papers see ref.~\cite{burdman}).

\subsection{Form factors}

A frequently encountered decomposition of the matrix elements relevant to the leptonic, 
the semileptonic and the penguin-induced hadronic matrix elements is
\bea\label{par1}
&&\langle 0 \vert \bar q \gamma_\mu \gamma_5 b \vert B(p_B)\rangle = i f_B p_{B\mu}\;,\cr
&&\langle P (p)\vert \bar q\gamma_\mu b \vert B(p_B)\rangle = 
\left( (p_B + p)_\mu - q_\mu {m_B^2 -
m_P^2\over q^2} \right)  F_+(q^2) + {m_B^2 - m_P^2 \over q^2} q_\mu
F_0(q^2)\;,\cr
&&\langle P(p) \vert \bar q\sigma_{\mu \nu}q^\nu b \vert B(p_B)\rangle \ =\
i\ \biggl( q^2 (p_B + p)_\mu - (m_B^2
- m_P^2) q_\mu \biggr) \ {\; F_T(q^2)\; \over m_B + m_P}\;,
\eea
where $q=d$ or $s$, $\vert P(p)\rangle$ is the light 
pseudoscalar meson state (pion or kaon), and $q^\nu=(p_B-p)^\nu$.

We will be working in the static limit, $m_B\to \infty$.  
The eigenstates of the QCD and HQET Lagrangians are  related as
\bea \label{norm}
\lim_{m_B\to \infty} {1\over \sqrt{m_B}} \vert B(p_B)\rangle_{\rm QCD} =  \vert
B(v)\rangle_{\rm HQET}\;.
\eea
In the static limit it is  more convenient to use definitions in which the form factors are 
independent of the heavy meson mass, namely, 
\bea \label{par2}
&&\langle 0 \vert \bar q  \gamma_\mu \gamma_5 b_v \vert B(v)\rangle_{\rm{ HQET}} = i \hat f v_\mu\ ,\cr
&&\cr
&&\langle P (p)\vert \bar q \gamma_\mu b_v \vert B(v)\rangle_{\rm{ HQET}} = 
\bigl[ p_\mu - (v\negcdot p) v_\mu\bigr] f_p (v\negcdot p)\ +\ v_\mu  f_v (v\negcdot
p)\;,
\eea
where the field $b_v$ does not depend on the heavy quark mass~\cite{Neubert:1993mb}. 
The form factors $f_{p,v}$ are functions of the variable 
\bea
v\negcdot p = { m_B^2 + m_P^2 - q^2 \over 2 m_B}\;,
\eea
which in the heavy meson rest frame is the energy of the light meson $E_P$. 
The relation between the quantities defined in eqs.~(\ref{par1}) and (\ref{par2}) 
is obtained by  matching QCD to HQET at the scale $\mu\sim m_b$~\cite{hill}:
\beq \label{match}
\begin{split}
f_B &= {C_{\gamma_0\gamma_5} (m_b) \over \sqrt{m_B}} \left( \hat f + {\cal O}(1/m_B) \right) \;,\\
 F_+(q^2) + {m_B^2 - m_P^2 \over q^2} &\left.\bigl[ F_+(q^2) - F_0(q^2) \bigr] \right|_{q^2 \approx
q^2_{\rm max}} \\
&= C_{\gamma_1} (m_b) \sqrt{m_B}
\left[ f_p(v\negcdot p) + {\cal O}(1/m_B)\right] \;,\\
(m_B + E_P) F_+(q^2) - 
(m_B &- E_P) {m_B^2 - m_P^2 \over q^2}\left.\bigl[ F_+(q^2) - F_0(q^2) \bigr] \right|_{q^2 \approx q^2_{\rm max}} \\
&=C_{\gamma_0} (m_b) \sqrt{m_B} f_v(E_P)  + {\cal O}(1/m_B)\;,\\
{2 m_B  \over m_B + m_P} F_T(q^2) &= C_{\gamma_1\gamma_0}(m_b) 
f_p(v\negcdot p)\sqrt{m_B} + {\cal O}(1/m_B)\;.
\end{split}
\eeq
In the following, we set the matching constants $C_{\gamma_i}$ to their tree level value ($C_{\gamma}=1$). 
By neglecting the subleading terms in the heavy quark expansion, we have~\footnote{Corrections at
${\cal O}(1/m_b)$ have been discussed in~\cite{boyd,nir,pirjol}.} 
\bea
\left.  F_0(q^2) \right|_{q^2 \approx
q^2_{\rm max}} =  {1\over \sqrt{m_B}}  f_v(v\negcdot p)\;, \quad \left.F_+(q^2) \right|_{q^2 \approx
q^2_{\rm max}}=\left.F_T(q^2) \right|_{q^2 \approx
q^2_{\rm max}}={\sqrt{m_B} \over 2} f_p(v\negcdot p)\,,
\eea
which exhibit the usual heavy mass scaling 
laws for the semileptonic form factors~\cite{isgur}. 
The equality $F_T(q^2)\equiv F_+(q^2)$ arises from 
\bea\label{TV}
\langle P(p)\vert \bar{q}\sigma^{i0} (1+\gamma_5)b_v \vert B(v)\rangle_{\text{HQET}} = i 
\langle P(p)\vert\bar{q}\gamma^{i} (1-\gamma_5)b_v\vert B(v)\rangle_{\text{HQET}} \,,
\eea
which is easily obtained by using the equation of motion $\gamma_0 b_v = b_v$.
In the following our main concern will be the evaluation of the long-distance 
(chiral) corrections to the form factors $f_{v,p}(v\negcdot p)$.

\subsection{Heavy-light current}

Having in mind eq.~(\ref{TV}), we only need to consider the ($V-A$) Dirac structure. 
In the static heavy quark limit and at next-to-leading order in the chiral expansion,  
the bosonized {\it heavy} $\to$ {\it light} current reads~\cite{booth}
\bea
\begin{split}
\label{current}
 J^\mu\equiv \bar{q}_a\gamma^\mu(1-\gamma_5)Q &\to \frac{i \alpha}{2}\Tr[\gamma^\mu (1-\gamma_5) H_b] \xi_{ba}^\dagger 
[1+V_L'(0)\Phi_0] \\
&+\frac{i \alpha}{2}\varkappa_1\Tr[\gamma^\mu (1-\gamma_5) H_c ]\xi_{ba}^\dagger ({\cal M_+})_{cb}\\
&+\frac{i 
\alpha}{2}\varkappa_2\Tr[\gamma^\mu (1-\gamma_5) H_b] \xi_{ba}^\dagger \str_c({\cal M_+})_{cc}\;.
\end{split}
\eea
When bosonizing the current $J^\mu$, one could envisage an arbitrary function 
$V_L(\Phi_0)$ in the first term on the right of eq.~(\ref{current}), expanded in a Taylor series.  
Only the terms linear in $\Phi_0$ are relevant to our purpose  and $V_L(0)$ 
can be normalized to $1$~\cite{zhang}. The appearance of $V_L^\prime(0)$ is yet 
another artifact of the quenched theory. The phase of the heavy meson can be chosen 
in such a way that $V_L^\prime(0)$ is completely imaginary, whereas the constants $\alpha$,
$\varkappa_1$, and $\varkappa_2$ are real. Notice that, at leading order in the chiral expansion, 
the constant $\alpha$ is simply the heavy-light meson decay constant in the static limit 
($m_Q\to \infty$), i.e.,  $\hat f$ in eq.~\eqref{par2}.

\section{Chiral Corrections to $f_p(v\negcdot p)$ and $f_v(v\negcdot p)$ \label{sec:chi-results}}

In this section we explain the main steps undertaken to compute the 
chiral logarithmic corrections for $\hat f$, $f_p(v\negcdot p)$, and 
$f_v(v\negcdot p)$. 
Our results for the decay constant $\hat f$  agree with those of refs.~\cite{booth,zhang}.
They will be used in Sec.~\ref{errors-ratios} to construct the ratios that are independent 
of counterterms. In ref.~\cite{falk} ChPT was applied to compute the heavy-to-light form factors. 
We repeat that calculation and extend it to the quenched case.

\subsection{$f_p^{\rm Tree}(v\negcdot p)$ and $f_v^{\rm Tree}(v\negcdot p)$ }

\begin{figure}
\begin{center}
\epsfig{file=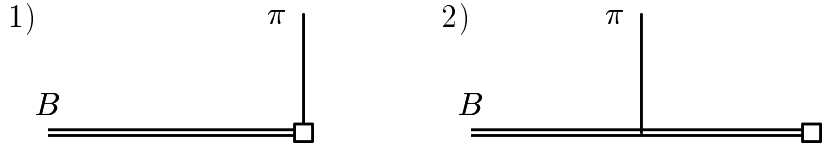, height=2.cm}
\caption{\label{fig1}\footnotesize{The point (1) and the pole (2) 
tree level Feynman diagrams contributing to 
$\text{\it heavy} \to \text{\it light}$ 
transition form factors. The box denotes the weak current insertion.}} 
\end{center}
\end{figure}
We start the discussion by writing the tree 
level expressions   
\bea
f_p (v\negcdot p) = {\alpha \over f} {g \over v\negcdot p + \Delta_i^\ast} \;, 
\qquad f_v (v\negcdot p) = {\alpha \over f} \;,
\qquad \hat f  = \alpha \;, 
\eea
The point and the pole diagrams, depicted in fig.~\ref{fig1}, describe $f_{v}$ and $f_{p}$, 
respectively. The index $``i"$ in $\Delta_i^\ast =m_{B_i^*}-m_{B_i}$  labels the light quark flavor. 
When necessary, we will use $P_{ij}$ to denote the light pseudoscalar mesons 
with valence quark content $ q_i\bar q_j$. 

Although the heavy quark spin symmetry suggests $\Delta_i^\ast \to 0$, we 
will keep this term finite because it provides the pole at $m_{B_i^*}^2$ to 
the form factor $f_p$, or equivalently to the form factors $F_{+,T}$.~\footnote{The pole dominance 
is easily seen if one rewrites the denominator of $f_p (v\negcdot p)$ as 
$v\negcdot p + \Delta^\ast = ({m_{B^\ast}/ 2}) \left( 1 - q^2/m_{B^\ast}^2 \right)$, 
where the corrections $(m_{B^*}-m_B)/m_B$ and $m_P^2/m_B^2$ are neglected.}
The form factor $f_v$ (or $F_0$), on the other hand, does not depend on  
$(v\negcdot p)$ at the tree level. 

\subsection{$f_p^{\rm one-loop}(v\negcdot p)$ and $f_v^{\rm one-loop}(v\negcdot p)$ }

\begin{figure}[h!] 
\begin{center}
\epsfig{file=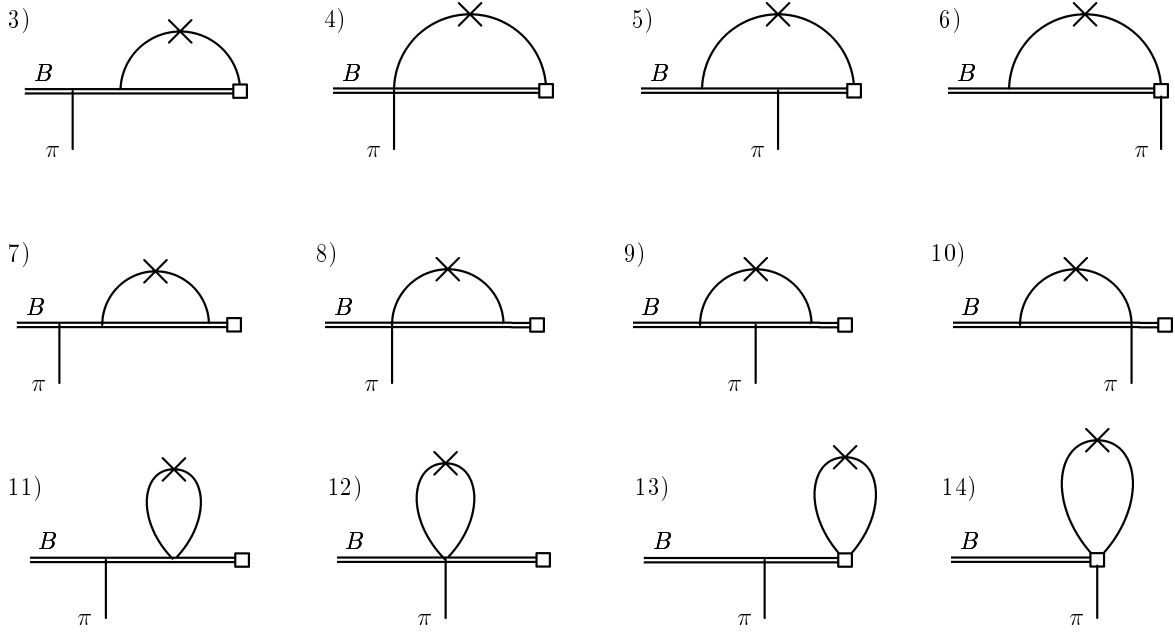, height=8.5cm}
\caption{\label{fig2}\footnotesize{The one-loop  contributions to 
the $B\to \pi $ transition. Double (single) lines denote the heavy (light) meson, 
while the weak current insertion is depicted by the empty box.  
Crosses stand for the $m_0$ hairpin vertex~\eqref{light}. 
Each graph represents two Feynman diagrams in quenched ChPT: 
($a$) the diagram  without the hairpin vertex (cross) and ($b$) 
the same diagram with a cross. In full ChPT only the diagrams 
without the cross are present. The amplitudes corresponding to the diagrams 
are listed in appendix~\ref{appB}. }} 
\end{center}
\end{figure}
Neglecting the crosses, all the graphs shown  in fig.~\ref{fig2} describe the one-loop chiral 
contributions to the form factors $f_{v,p}(v\negcdot p)$ in the full ChPT.
In the quenched theory, in contrast, graphs both with and without crosses appear. 
The cross denotes the so-called hairpin vertex, i.e., the dipole term in the ``pion" propagator: 
\bea
{1\over p^2 - m_P^2} \left( \begin{array}{cc}   1& 0\\ 0 & -1 \end{array}\right) -
{\alpha_0 p^2 - m_0^2 \over (p^2 - m_P^2)^2}\left( \begin{array}{cc}   1& 1\\ 1 & 1 \end{array}\right)\;.
\eea

In the computation of the loop integrals, 
the naive dimensional regularisation has been used, with the renormalization prescription  
$\msbar +1$,  where  $\bar \Delta= 2/\epsilon-\gamma+\ln(4\pi)+1$ is subtracted~\cite{gasser}. 
We neglect the isospin symmetry breaking ($m_u = m_d\equiv m_q$), as well as the mass 
differences between $B$, $B_s$, $B^\ast$, $B_s^\ast$ meson states, whenever they 
appear in the loop. The resulting expressions can be written as 
\bea \label{QCHPT}
&& \centerline{\underline{\sf \; Quenched\ ChPT\; }}  \nonumber \\
&&  \nonumber \\
&& f_p^{B_j\to P_{ij}}(v\negcdot p) = 
                 {\alpha \over f}\ {g \over v\negcdot p + \Delta^\ast_i} \ 
		\left[ 1 + \delta f_p^{B_j\to P_{ij}} + {k_1\over 2} m_j + \kappa_1 m_i- 
		  4L_5  {4 \mu_0 \over f^2}  (m_i+m_j)\right]\;, \nonumber \\
&& f_v^{B_j\to P_{ij}}(v\negcdot p) = {\alpha \over f}\   
		\left[ 1+  \delta f_v^{B_j\to P_{ij}} + 
		\left( {k_1\over 2} +\varkappa_1\right) m_j 
	        -4L_5{4 \mu_0 \over f^2}  (m_i+m_j) \right]\;,   \nonumber \\
&& \hat f_j =  \alpha \ \left[ 1 + \delta \hat f_j  + \left( {k_1\over 2} + 
		\varkappa_1\right) m_j\right]\;, 
\eea

\bea \label{SCHPT}
&& \centerline{\underline{\sf \; Full\ ChPT\; }}  \nonumber \\
&&  \nonumber \\
&& f_p^{B_j\to P_{ij}}(v\negcdot p) = 
                 {\alpha \over f}\ {g \over v\negcdot p + \Delta^\ast_i} \ 
		\left[ 1 + \delta f_p^{B_j\to P_{ij}} + {k_1\over 2} m_j 
		+\kappa_1 m_i- 
		   4L_5 {4 \mu_0 \over f^2}  (m_i+m_j)\right.  \nonumber \\
&&  \left. \hspace*{70mm} + \left( {k_2 \over 2} - 
8L_4 {4 \mu_0 \over f^2} +\kappa_2\right) (m_u+m_d+m_s) \right]\;, \nonumber \\
&& f_v^{B_j\to P_{ij}}(v\negcdot p) = {\alpha \over f}\   
		\left[ 1+  \delta f_v^{B_j\to P_{ij}} + 
		\left( {k_1\over 2} +\varkappa_1\right) m_j 
	        -4L_5 { 4 \mu_0 \over f^2}  (m_i+m_j) \right.   \nonumber \\
&&  \left. \hspace*{70mm} + \left( {k_2 \over 2}+\varkappa_2 - 8L_4 {4 \mu_0 \over f^2} \right) (m_u+m_d+m_s) \right]\;, \nonumber \\
&& \hat f_{j} =  \alpha \ \left[ 1 + \delta \hat f_{j}  + \left( {k_1\over 2} + 
		\varkappa_1\right) m_j + \left( {k_2\over 2} +\varkappa_2\right) (m_u+m_d+m_s) \right]\;. 
\eea
We separated the chiral loop corrections 
($\delta f$) from those involving the counter-terms. The loop corrections to the form factors  are written as
\bea
\delta f_{v,p}^{B_j\to P_{ij}} = \sum_I \delta f_{v,p}^{(I)} + {1 \over 2} \delta Z_{B_j}^{\text{loop}} +  {1 \over
2}\delta Z_{P_{ij}}^{\text{loop}}\;,
\eea
where the sum runs over all diagrams shown in fig.~\ref{fig2}, and 
$\delta Z_{B_j (P_{ij})}$  encodes the chiral loop contributions arising from 
the wave function renormalization of the heavy--light and light-light meson respectively.
The explicit expressions are lengthy and are collected in appendixes~\ref{appBQuenched} (for the quenched) 
and~\ref{appBUnQuenched} (for the full theory). 
As can be seen from eqs.~(\ref{QCHPT}) and~(\ref{SCHPT}), no dependence on $(v\negcdot p)$ arises 
from the counterterms. The  modification of the tree-level 
$(v\negcdot p)$ dependence of the form factors $f_{v,p}$ is entirely due to the chiral loop corrections.
It is worth noticing that in the quenched approximation the form factor 
$f_v^{B\to \pi}$ is completely independent of $(v\cdot p)$. This is so because the contribution from 
 diagram 4 in fig.~\ref{fig2}  vanishes in the isospin limit $m_u=m_d$.

An important feature emerging from this calculation is that the quenched form factors $f_{v,p}$ and quenched decay 
constant $\hat f$ suffer from the common  quenched pathology, that is, from the presence of the quenched chiral logs  
$\propto m_0^2 \log (m_P^2/\mu^2)$. Such terms are divergent in the limit $m_P\to 0$,  suggesting that 
in the quenched approximation the chiral limit for any of $F^{B\to \pi}_{+,0,T}$ or $f_B$ is not defined. 
This feature is also present in the light meson sector, e.g., for the  chiral condensate, the light meson 
decay constant consisting of nondegenerate quark flavors, etc.~\cite{golterman,sharpe-review}.

\section{Choice of Parameters\label{sec:parameters}}
For numerical analysis of the expressions given in eqs.~\eqref{QCHPT} and (\ref{SCHPT}) 
we need to make a specific choice of quite a number of low energy
constants. 
To do so we will mainly rely on the existing lattice data. In table~\ref{tab1} we collect the parameters whose (range
of) values we were able to fix. In the following we briefly discuss each of those values.
\begin{table}[h]
\centering 
\begin{tabular}{ccc}  \hline \hline
{\phantom{\huge{l}}}\raisebox{-.2cm}{\phantom{\Huge{j}}}
{\sl Coupling} & {\sf Full theory}&  {\sf Quenched theory}   \\ 
\hline 
{\phantom{\huge{l}}}\raisebox{-.2cm}{\phantom{\Huge{j}}}
$\alpha$ ($\gev^{3/2}$)  & $0.56 \pm 0.04$  & $0.48 \pm 0.03$  \\ 
{\phantom{\huge{l}}}\raisebox{-.2cm}{\phantom{\Huge{j}}}
 $g$  & $0.50 \pm 0.10$  & $0.56 \pm 0.12$  \\   
{\phantom{\huge{l}}}\raisebox{-.2cm}{\phantom{\Huge{j}}}
$f$ ($\gev $)  & $0.130$  & $0.124  \pm 0.004$  \\ 
{\phantom{\huge{l}}}\raisebox{-.2cm}{\phantom{\Huge{j}}}
$\mu_0$ ($\gev $)  & $1.14  \pm 0.20$  & $1.13  \pm 0.04$  \\ 
{\phantom{\huge{l}}}\raisebox{-.2cm}{\phantom{\Huge{j}}}
$L_4$ ($\times 10^{-3}$)  & $-(0.5 \pm 0.5)$   & $(0.0 \pm 0.5)$  \\ 
{\phantom{\huge{l}}}\raisebox{-.2cm}{\phantom{\Huge{j}}}
$L_5$ ($\times 10^{-3}$)  & $0.8 \pm 0.5$   & $0.8 \pm 0.2$  \\ 
{\phantom{\huge{l}}}\raisebox{-.2cm}{\phantom{\Huge{j}}}
$m_0$ ($\gev $)  & --  & $0.64 \pm 0.06$  \\ 
{\phantom{\huge{l}}}\raisebox{-.2cm}{\phantom{\Huge{j}}}
$g^\prime$ ($\gev $)  & --  & $-0.6 \; {\text{to}}\; 0.6$  \\ 
\hline
\hline
\end{tabular}
{\caption{\small \label{tab1}Low energy constants whose determination is discussed in the
text.}}
\end{table}

\begin{itemize}
\item[$\circ$] {\underline{$\boldmath{ \alpha}$}}: This constant is equal to 
the heavy-light decay constant in the static limit of QCD. Its value can be obtained 
from $f_D\sqrt{m_D}$ and/or $f_B \sqrt{m_B}$, which are then extrapolated (in inverse 
heavy-light meson mass) to the infinite mass limit as
\bea \label{eq3}
f_H \sqrt{m_H} = \alpha \left( 1 + A/m_H + B/m_H^2 + \dots\right)\;.
\eea 
From lattice QCD and QCD sum rules, it is known that the slope $A$ is large and 
negative, whereas the value of the curvature $B$ is small. Therefore, to a good accuracy, 
one can set $B=0$ and neglect higher terms in $1/m_H$. In this way we obtain the following:
\begin{itemize}

\item From extensive quenched lattice simulations by the MILC collaboration, one can deduce 
$\alpha^{\rm quench} = 0.45(5)\ \gev^{3/2}$~\cite{milc}. 
In the same paper, they also present the results of their unquenched simulations, from which 
we extract $\alpha^{\rm full} = 0.53(7)\ \gev^{3/2}$. These numbers agree quite well with the 
values obtained by CP-PACS Collaboration, namely, $\alpha^{\rm quench} = 0.50(4)\ 
\gev^{3/2}$, and $\alpha^{\rm full} = 0.57(6)\ \gev^{3/2}$~\cite{cppacs}. 
Notice that both references use the same treatment of the heavy quark on the lattice 
(the so-called Fermilab formalism).

\item The UKQCD~\cite{lacagnina} and APE~\cite{ape2} Collaborations compute heavy-light 
meson decay constants using the fully relativistic lattice QCD, but only for the mesons of 
masses $m_H \in (1.8, 2.7)$~GeV. 
From a linear fit of the form~(\ref{eq3}), from their quenched data UKQCD 
obtain~\footnote{We thank G.~Lacagnina for communicating this result to us.} 
$\alpha^{\rm quench}=0.49(4)~\gev^{3/2}$, in agreement with the previous result 
by APE, $\alpha^{\rm quench}=0.48(5)~\gev^{3/2}$.

\item Recent results obtained by using the QCD sum rules agree quite well with the above 
unquenched values. From the compilation of the QCD sum rule estimates in ref.~\cite{narison}, 
one has $\alpha^{\rm full} = 0.58(9)\ \gev^{3/2}$ (for the most recent result see~\cite{steihauser}).
\end{itemize}

\item[$\circ$] {\underline{$\boldmath{g}$}}: This constant is related to the phenomenological 
coupling $g_{D^\ast D\pi}$ as 
\bea
g_{D^\ast D\pi} = {2 \sqrt{m_D m_{D^\ast}}\over f_\pi} g\;, 
\eea
where one can also set $m_{D^\ast}= m_D$ because the above relation is defined only in the 
static limit, in which the heavy quark spin symmetry is exact.  From the experimentally 
measured total width of the $D^{\ast +}$ meson, one gets $g_c^{\rm full} = 0.59(7)$~\cite{cleo}. 
The subscript $``c"$ warns us that the value is obtained in the charm quark mass sector. 
On the lattice, that value has been computed recently in the quenched approximation, 
leading to $g_c^{\rm quench} = 0.66(9)$~\cite{spqr}. To get the value of $g$ one needs 
to extrapolate to the infinite heavy quark mass limit, i.e., $g_Q = g + \gamma/m_H + \dots$ 
The lattice data of  ref.~\cite{spqr} suggest that the slope $\gamma$ is negligible, and thus 
$g/g_c\approx 1$. On the other hand, the LCSR calculation suggests that the same slope 
is significant and positive~\cite{khodj}. By neglecting the terms of ${\cal O}(1/m_H^2)$  and higher, 
ref.~\cite{khodj} leads to $g/g_c \approx 0.7$. We will take the average of the two 
(lattice and LCSR), that is,  $g/g_c \approx 0.85$, and add the difference 
in the error bars of both quenched and unquenched values.~\footnote{We are aware of the 
result of ref.~\cite{ukg}, $g=0.42(9)$, where this coupling was computed for the first 
time on the lattice, in static HQET. However, in view of the insignificant statistics, 
very coarse lattice, and only two light quark masses, the value $g=0.42(9)$ should be considered 
as exploratory. An improved calculation of $g$, along the lines described in ref.~\cite{ukg} 
would be most welcome, though.}

\item[$\circ$] {\underline{$\boldmath  f$}}: To get the chiral coupling constant we will use  
$f_\pi = 132$~MeV and $f_K = 160$~MeV, and the fact that $m_s/m_q = 24.4\pm 1.5$~\cite{leutwyler}. 
After linearly extrapolating to the chiral limit ($m_q\to 0$), we get $f = 130$~MeV. 
The recent extensive quenched lattice study with Wilson fermions gives 
$f=119\pm 7$~MeV~\cite{aoki1}, whereas the one with staggered fermions gave  
$f=125\pm 9$~MeV~\cite{aoki2}. The latter result was also obtained on  
larger lattices with domain wall fermions, namely, $f=125\pm 7$~MeV~\cite{dwf1}. 
As a weighted average, we will take $f = 124 \pm 4$~MeV.

\item[$\circ$] {\underline{$\boldmath \mu_0$}}: 
From the Gell-Mann--Oakes--Renner formula \cite{Gell-Mann:rz}, it is easy to identify 
$\mu_0 = -\langle \bar q q\rangle/f_\pi^2$. Its value (in the full theory) can be fixed 
by using $\langle \bar q q\rangle^\msbar (2 \ \gev) = -[ 267(16)~\mev]^3$~\cite{jamin}. 
The quenched value is extracted from the lattice data. Recent results in the $\msbar$ 
scheme (at the renormalization scale $2$~GeV) are $\mu_0 = 1.13(4)~\gev$~\cite{aoki1}  
and $\mu_0 = 1.10(11)~\gev$~\cite{spqr-M}. With these numbers and by using $f = 124 \pm 4$~MeV, 
the corresponding numerical values for the chiral condensate in the quenched approximation are 
$\langle \bar q q\rangle^\msbar (2 \ \gev) = -[ 259(6)~\mev ]^3$ and 
$\langle \bar q q\rangle^\msbar (2 \ \gev) = -[ 257$ 
$(10)~\mev ]^3$.

\item[$\circ$] {\underline{$\boldmath L_{4,5}$}}: These two couplings have been extensively 
discussed  in the literature~\cite{chiral-reviews}. 
Their values in the full theory, at the scale $\mu \simeq 1$~GeV, are given in table~\ref{tab1}. 
In quenched QCD, the coupling $L_5$ was recently computed in ref.~\cite{alpha}, 
with the result  $\alpha_5 = 0.99(26)$ and $L_5 = \alpha_5/(128 \pi^2)=0.8(2)\times 10^{-3}$.  On the other hand, the quenched estimate of $L_4$ is 
not available. On the basis of the large $N_c$ expansion  we only know that 
$L_4$ is smaller than $L_5$ (see ref.~\cite{gasser}). We will take it to be zero 
and vary by $\pm 0.5 \times 10^{-3}$.

\item[$\circ$] {\underline{$\boldmath m_0$}}: This mass characterizes the 
magnitude of the hairpin insertion (crosses in fig.~\ref{fig2}). It enters in the 
coefficient of the quenched chiral log, which, in the literature, is often referred to as  
$\delta = m_0^2/(24 \pi^2 f_\pi^2)$. 
The precise value of $\delta$ is unknown at present, mainly because in realistic lattice 
studies it is very difficult (if possible at all) to resolve the finite lattice volume 
effects from the quenched chiral logs. This is why it is still not completely clear 
whether the observed deviations from a linear dependence of the pseudoscalar meson 
squared ($m_P^2$) on the light quark masses ($m_q$) is due to the presence of the 
chiral logs, or if it is an artifact of the lattice geometry (see, e.g., ref.~\cite{durr}). 
With this remark in mind, we now quote recent values for the parameter $\delta$, 
as obtained from lattice studies of the dependence of $m_P^2$ on $m_q$:
\begin{itemize}
\item with Wilson fermions, $\delta = 0.10(2)$~\cite{aoki1};
\item  with modified Wilson fermions,  $\delta = 0.073(20)$~\cite{eichten};~\footnote{
The final result of ref.~\cite{eichten}, $\delta = 0.065(13)$, is obtained by 
combining various ways of extracting this quantity from the lattice data. For an easier 
comparison with other groups, we quote the result given in eq.~(29) of ref.~\cite{eichten}, 
which is obviously fully consistent with their final value.}
\item with domain wall fermions, $\delta = 0.029(7)$~\cite{dwf2};
\item with overlap fermions, $\delta = (0.0$--$0.2)$~\cite{degrand} 
and $\delta = (0.2$--$0.3)$~\cite{kentucky}.
\end{itemize}
For the pre-1996 results see the review in~\cite{sharpe-review}.  
The value of $\delta$ obtained by the CP-PACS Collaboration stands out because 
they made an extensive high statistics study on  large lattice volumes and 
extrapolated to the continuum~\cite{aoki1}. A worrisome aspect of this value, however, 
is that the light pseudoscalar mesons that they access directly lie in the 
range $m_P\in (300, 750)$~MeV, 
for which  the significance of the chiral logs may be questionable. Keeping this  
comment in mind and using their $\delta = 0.10(2)$, we get $m_0 = 0.64(6)$~GeV.

\item[$\circ$] {\underline{$\boldmath g^\prime$}}: The value of this constant can be obtained from 
$g_{B B^\ast \eta^\prime}$ and $g_{D D^\ast \eta^\prime}$, the couplings of 
the lowest lying heavy-light meson doublet with the light quenched $\eta^\prime$ state. 
Such a lattice study has not been made so far. To get a rough estimate we may 
rely on the large $N_c$ limit from which one expects $\vert g^\prime\vert < g$.

\end{itemize}

\noindent
As for the other parameters appearing in the quenched expressions, in the numerical analysis,
we will first set them to zero and then vary their values in the ranges suggested by the large 
$N_c$ expansion. More specifically, we take  $\vert V_L^\prime\vert \leq 0.5$, 
$\vert \varkappa_1\vert ,\ \vert k_1\vert \leq 32 \mu_0 L_5/f^2$, and 
$\vert \varkappa_2\vert = \vert k_2\vert = 0$. The parameter $\alpha_0$ will
be varied in the range $\vert \alpha_0\vert < 0.1$. It is important to stress that the effect of the 
variation of these latter parameters is completely negligible for our numerical estimates.

Note that the low energy constants in table \ref{tab1} are extracted from lattice data at the 
tree level of ChPT (except for the counterterms).~\footnote{ 
The standard 
procedure would involve ``undressing" the chiral loops 
 from the measured low energy constants. 
 This procedure is not applicable when using the lattice data because
 direct lattice computations are made at $m_\pi \gtrsim 400$~MeV, 
and the physical results are obtained through a linear or quadratic 
extrapolation in $m_\pi^2 \to (m_\pi^2)^{phys}$. 
These extrapolations do not include the chiral logs, 
so ``undressing"  the chiral loops would lead to  unrealistic 
estimates of the low energy constants. For this reason, we adopted the tree-level approximation 
in extracting the low energy constants. }

\section{Quenching Errors\label{sec:numerics1}}

In this section we will use the expressions for the form 
factors~(\ref{QCHPT}) and (\ref{SCHPT}), the parameters listed in table~\ref{tab1},
to get an estimate of the quenching errors. We reiterate that the results of 
this section refer to the zeroth order in the $1/m_H$ expansion and to the leading 
logarithmic chiral corrections, with the specific choice of parameters discussed 
in the previous section.  We also stress that in all the following discussion 
we will keep the strange quark mass fixed to its quenched value of  
$m_s^{\msbar} (2\ \gev) = 105$~MeV, and $m_s^{full}/m_s^{quench} \simeq 0.85$~\cite{wittig}.
 To examine the quenching errors we will study the following ratios 
\bea \label{Q-ratios}
Q_{p,v}(v\negcdot p, m_q/m_s) = {f_{p,v}^{full}(v\negcdot p) -  
f_{p,v}^{quench}(v\negcdot p)\over f_{p,v}^{full}(v\negcdot p)} \,.
\eea
The parameter that makes the strongest impact on the results for 
$Q_{p,v}$ is $ g^\prime$. As stated above, its value is expected to lie in the range  
$-g \leq g^\prime \leq g$. Since even its sign is not known  
we will distinguish between  the two ``extreme" cases $g^\prime = -g$ and $g^\prime = +g$.

\subsection{$B\to \pi$ transition}
\begin{figure}
\begin{center}
\hspace*{-8mm}\epsfig{file=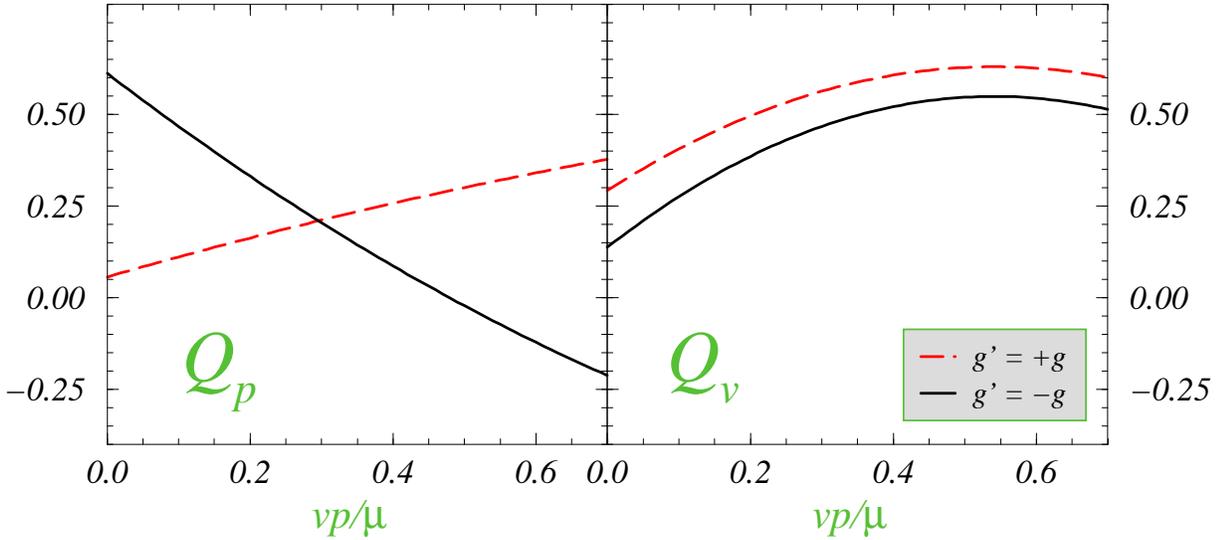,width=\linewidth}
\caption{ \label{plot1}\footnotesize{The ratios $Q_{p,v}(v\negcdot p)$ are defined in eq.~(\ref{Q-ratios}). 
The plots refer to the mesons consisting of two degenerate (valence) quarks with mass 
$m_q=r~m_s$ and  $r= 0.25$. The counter-term coefficients $k_{1,2}$, $\varkappa_{1,2}$, as well as 
$\vert V_L^\prime (0)\vert$ are neglected, while $\mu  = 1~\text{GeV}\simeq \Lambda_\chi$. 
Values of other parameters are given in table \ref{tab1}.
}} 
\end{center}
\end{figure}
To examine the quenching errors for $B\to \pi$ decay  we cannot set the pion mass  
to its physical value, because the spurious quenched chiral logs would become dominant. 
However, we cannot go far away from the chiral limit either, because the (Q)ChPT approach 
then becomes inadequate. To those two competing requirements, we should add a third:
a desire to be sufficiently close to the region of quark masses (i.e., pseudoscalar meson
masses squared) probed by the current quenched lattice studies~\cite{ukqcd,ape,jlqcd,fnal,shigemitsu}. 
It is not clear whether or not a mass interval for which all of the above requirements 
are satisfied exists. We will assume that they are satisfied for $m_P\approx 330$~MeV. 
From eq.~(\ref{gell-mann}), a ``pion" of mass $m_P\approx 330$~MeV is composed of two quarks of 
mass corresponding to $r=m_q/m_s\simeq 0.25$, with respect to the strange quark mass $m_s$.
In fig.~\ref{plot1} we plot the ratios $Q_{p,v}(v\negcdot p, r=0.25)$.

We notice that, regardless of the value chosen for $g^\prime$, 
the quenching errors in the form factor $f_p(v\negcdot p)$ are not excessively 
large for  most of the $v\negcdot p$ where ChPT is applicable. This is 
important since this form factor is the only one entering the $\bpi$ decay rate, from which 
we hope to be able to extract the value for $\vert V_{ub}\vert$. 
  
\begin{figure} 
\begin{center}
\hspace*{-8mm}\epsfig{file=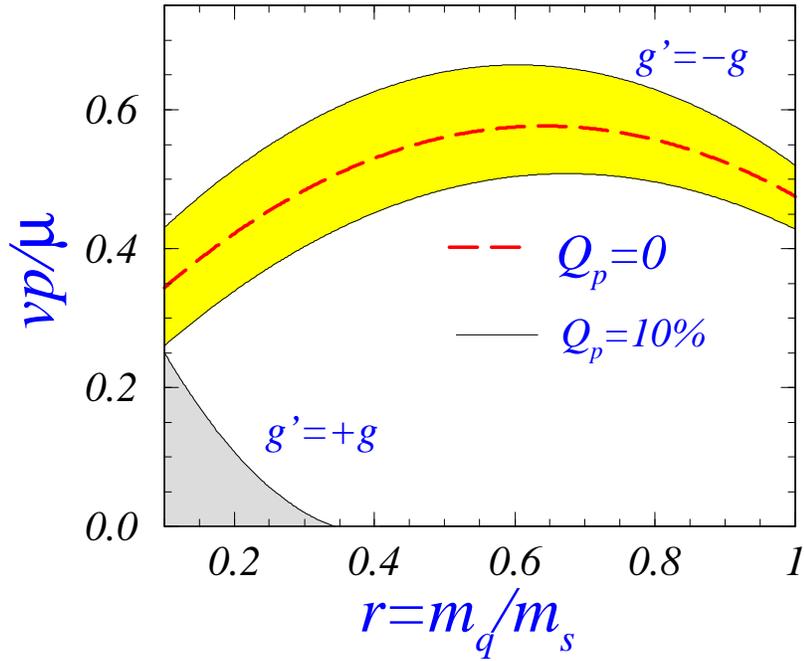,width=.66\linewidth}
\caption{\label{plot2}\footnotesize{The thick dashed line corresponds 
to the zero quenching errors [$Q_p(v\negcdot p,r)=0$] in the case 
when $g^\prime=-g$ (such a line is not accessible for $g^\prime=+g$). 
The shaded region corresponds to the variation of $Q_p(v\negcdot p,r)$ 
by $10\%$. Notice that in the case  $g^\prime=-g$, the upper 
(lower) curve corresponds to $Q_p(v\negcdot p,r)$ equal to $- 10\%$ 
($+ 10\%$). For the case $g^\prime=+g$, only the region  $Q_p(v\negcdot p,r) 
<+10\%$ is shown.}} 
\end{center}
\end{figure}

By specifying $g^\prime=+g$, we observe that $Q_p(v\negcdot p) >0$, for any $ (v\negcdot p)/\mu < 0.7$ and  $r\geq 0.2$.   
In other words, quenched values for $f_p$ are smaller than unquenched ones. If we take $g^\prime=-g$ instead, 
the ratio $Q_p(v\negcdot p)$ has a zero. The point of zero quenching 
errors in Fig. \ref{plot1} is found at $(v\negcdot p/ \mu) = 0.47(3)$ for 
$r=0.25$ and the values of parameters given in table \ref{tab1}. Figure~\ref{plot2} shows 
the curve of vanishing quenching errors in the $v\negcdot p-r$ plane, i.e., 
$Q_p(v\negcdot p,r)=0\pm 10\%
$. This curve really exists for $g^\prime=-g$, while for $g^\prime=+g$ only the part corresponding 
to $Q_p(v\negcdot p,r)= + 10\%
$ can be reached for $r<0.4$. In summary, it is possible to find combinations of the pion mass and the pion recoil energy 
such that the quenching errors in the dominant form factor $f_p(v\negcdot p)$ are 
kept under the $10\%
$ level.  

>From fig.~\ref{plot1} we also see that  
there exists a point $(v\negcdot p/\mu) \approx 0.3$ such that the ratio $Q_p$ is independent 
of the value of $g^\prime$. At that point, we get 
\bea \label{estimate}
Q_p(0.3 \mu ,r\simeq 0.25) \simeq 20\%\;. 
\eea
This result is a useful estimate of the 
quenching error for realistic values of $(v\negcdot p)$ and $r$, which are 
currently probed on the lattice.  For the reader's convenience, we have fitted 
the points corresponding to $(\partial f_p/\partial g^\prime)=0$ to a polynomial 
in $r\in [0.1,1]$ and obtained
\bea \label{nogprime}
{v\negcdot p  \over \mu} = 0.17(1) +0.8(1) r -1.3(2) r^2 + 0.58(2) r^3\;.
\eea
We also checked that for $0.1<r\lesssim 0.35$, with $r$ and $(v\negcdot p)$ 
satisfying eq.~(\ref{nogprime}), the quenching errors are kept under the $25\%
$ level.
This concludes our discussion of the form factor $f_p$.

The situation with the form factor $f_v(v\negcdot p)$ is much worse. From 
fig.~\ref{plot1}, we see that the quenching errors are in the range $30-60\%
$, and drop below that level only for larger recoils for which the present approach is not appropriate. 
The quenching errors remain large when varying $r$ in the range  $r \in [0.1,1]$. 
A somewhat less pessimistic situation is present at zero recoil ($v\negcdot p=0$) where, 
contrary to the case of $f_p$, the form factor $f_v$ {\it can} be extracted from the lattice data. 
For $m_P = 330$~MeV, at zero recoil we get
\[
Q_v(0,r=0.25) \simeq \left\{ \begin{array}{ccc}
 8\% & & (g' = +g)\,,\\
&& \\
 23\%  & & (g' = -g)\,.\\
\end{array} \right.
\]
Once again, we warn the reader that the numerical estimates made in this section 
are obtained for the set of low energy constants specified in table~\ref{tab1}.

\subsection{$B\to K$ transition}
We now turn to the case of a kaon in the final state. In the discussion
 we shall fix one of the valence quarks to 
the strange quark mass, and vary the other one ($m_q=r~m_s$) in the range  
$1/25 \lesssim r < 0.5$. In this way the ``kaon" mass is varied in the range $m_K^{phys}< m_P < 600$~MeV.
For simplicity we consider $(v\negcdot p)/\mu \simeq 0.3$  (reasonably small recoil), 
and examine the quenching ratios
\bea \label{QK-ratios}
Q^{K}_{p,v}(0.3 \mu, r) = {f_{p,v}^{full}(v\negcdot p) -  
f_{p,v}^{quench}(v\negcdot p) \over f_{p,v}^{full}(v\negcdot p)} \,.
\eea
\begin{figure} 
\begin{center}
\hspace*{-8mm}\epsfig{file=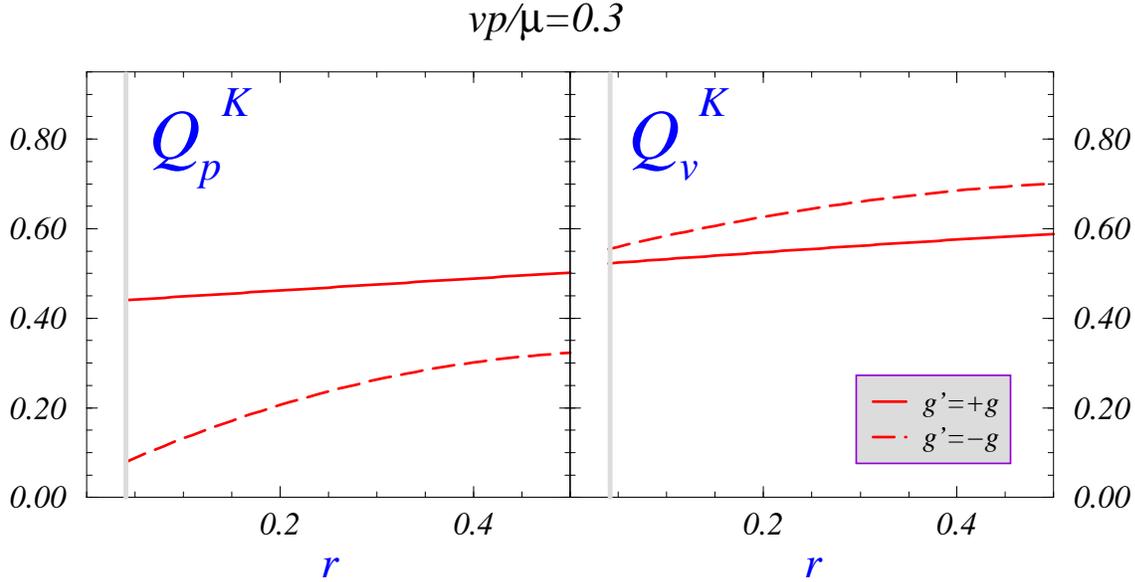,width=.93\linewidth}
\caption{\label{plot3}\footnotesize{The ratios $Q_{p,v}(v\negcdot p)$ from eq.~(\ref{QK-ratios}).  
The plots refer to mesons whose one valence quark is fixed to the $s$-quark mass, while 
the other has the value $m_q=r m_s$. The thick gray vertical line marks $r\simeq 1/25$ for which the 
physical kaon mass is reached. As before, the counterterm coefficients $k_{1,2}$, $\varkappa_{1,2}$, as well as 
$V_L^\prime (0)$ are neglected, while $\mu = 1$~GeV. 
}} 
\end{center}
\end{figure}
\noindent
In the quenched expressions for $f_{p,v}$, given in appendix~B, we set for the active quark 
$M_i^2 = 4 \mu_0 m_s$, while for the spectator one we take  $M_j^2 = 4 \mu_0 m_s r$ [see eqs.~\eqref{fpQ}, 
\eqref{fvQ}]. As in the previous subsection, in fig.~\ref{plot3} we plot $Q^{K}_{p,v}$ for the two ``extreme" 
scenarios, namely, $g^\prime = +g$ and $g^\prime = -g$.
We again observe that, regardless of the value 
for $g^\prime$, the quenching errors on the form factor $f_v(v\negcdot p)$,  $Q^{K}_{v} > 0.5$. 
The quenching errors $Q^{K}_{p}$ are only moderately large  
if $g^\prime =-g$, and are large for $g^\prime =+g$.

Therefore, from this and the preceding subsection,  we see that in both channels 
$B\to \pi$ and $B\to K$ our approach suggests that the quenching errors 
in the form factor $f_v(v\negcdot p)$ [i.e., $F_0(q^2)$] are large $\gtrsim 30$\%. 
The quenching errors in  $f_p(v\negcdot p)$ [i.e., $F_{+,T}(q^2)$], on the other hand,
depend crucially on the value of the low energy constant $g^\prime$: 
they are large for $g^\prime > 0$, and {\it ``not so large"} for $g^\prime < 0$.

Before closing this subsection let us  mention that in the  quenched  lattice 
studies the kaon is usually considered as a composite state of two degenerate 
quarks. Using $m_P^2 = 4\mu_0 m_s r$, one  varies $r$ in order to reach the physical kaon 
mass $m_P = m_K^{phys}$, which occurs for $r\simeq 0.5$. One may wonder 
if the form factors with such a kaon differ from the ones in which the quarks in 
the kaon are nondegenerate, with one of the quarks fixed to the strange mass and the other 
varying toward $r\simeq 1/25$ (i.e., the physical kaon mass). 
To keep the mass of the pseudoscalar meson the same in both situations, 
we will vary $r_{deg.} \in [0.6,0.8]$ and $r_{ndg.} = 2 r_{deg.}-1 \in [0.2,0.6]$. As before, 
we take $(v\negcdot p/ \mu) =0.3$ and examine the following ratio
\bea \label{deg}
R_{v,p}(r_{deg.}) = \left. {f_{v,p}^{ndg.}(v\negcdot p,r_{ndg.}) - 
f_{v,p}^{deg.}(v\negcdot p,r_{deg.}) \over f_{v,p}^{ndg.}(v\negcdot
p,r_{ndg.})}\right|_{r_{ndg.} = 2 r_{deg.}-1} \;
\eea
\begin{figure}
\begin{center}
\hspace*{-8mm}\epsfig{file=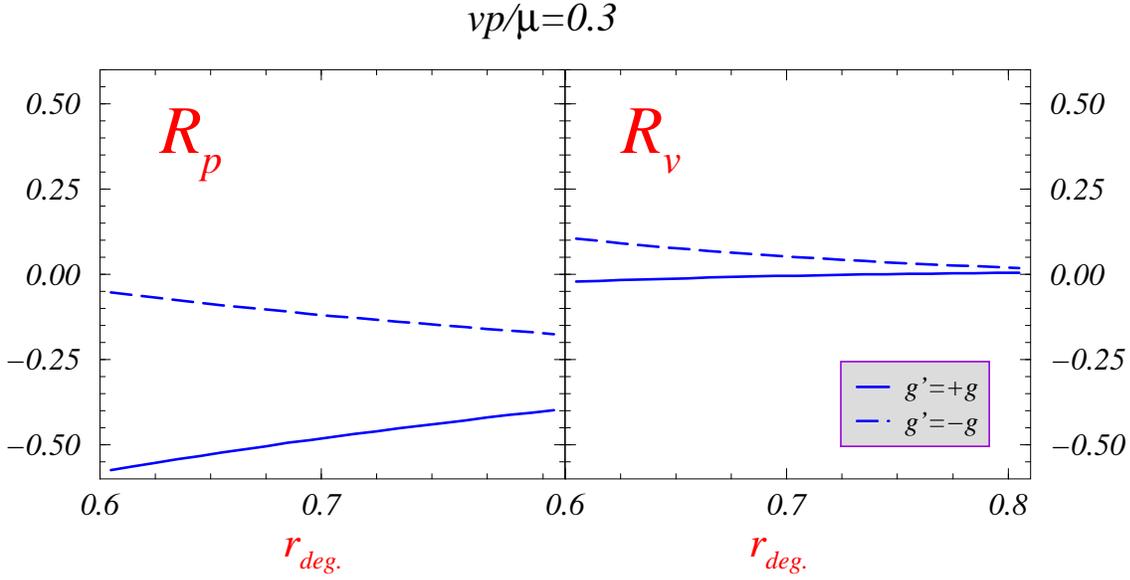,width=.93\linewidth}
\caption{\label{plot4}\footnotesize{The ratios $R_{p,v}$ defined in eq.~(\ref{deg}), 
measuring the errors induced by the degeneracy in the ``kaon" quark masses in 
the quenched calculations.  
}} 
\end{center}
\end{figure}
in the quenched theory. From fig.~\ref{plot4}, we again observe the same dichotomy: for $g^\prime =-g$ 
the situation is quite favorable, i.e., the errors due to degeneracy in 
the quark masses are very small, whereas for  $g^\prime =+g$ the form factor 
$f_p^{deg.}$ is highly overestimated with respect to $f_p^{ndg.}$.

\subsection{Useful ratios}\label{errors-ratios}

The double ratio $f_v^{B_s\to K} f_{B}/f_v^{B\to K} f_{B_s}$ is very gratifying 
from the ChPT point of view. In a double ratio of the form
\begin{equation}
\frac{f_{v(Q)}^{B_j\to P_{ij}}}{f_{v(Q)}^{B_i\to P_{ji}}}~
\frac{f_{B_i(Q)}}{f_{B_j(Q)}}=1+\delta f_{v(Q)}^{B_j\to P_{ij}}-\delta 
f_{v(Q)}^{B_i\to P_{ji}}+\delta f_{B_i(Q)}-\delta f_{B_j(Q)} \;,\label{eq:002}
\end{equation}
where $B_j\sim b \bar q_j$ and $P_{ij}\sim q_i \bar q_j$, the dependence on $\alpha$ 
cancels completely and so do the counterterms. Quenching errors in this quantity are 
thus far more reliably predictable in the framework of ChPT than for the separate form 
factors. This quantity might then prove useful in future lattice simulations with both 
$B_s\to P$  and $B\to P$ transitions.

A similar (partial) cancellation of dependence on counterterms occurs also for the quenching 
error in the ratio  $f_{p,v}^{B\to K}/f_{p,v}^{B\to \pi}$ 
\begin{equation}
\begin{split}
\frac{f_{p,v(Q)}^{B_j\to P_{ij}}}{f_{p,v(Q)}^{B_k\to P_{lk}}}-
&\frac{f_{p,v}^{B_j\to P_{ij}}}{f_{p,v}^{B_k\to P_{lk}}}=\\\delta 
&f_{p,v(Q)}^{B_j\to P_{ij}}-\delta f_{p,v(Q)}^{B_k\to P_{lk}}-\delta 
f_{p,v}^{B_j\to P_{ij}}+\delta f_{p,v}^{B_k\to P_{lk}}- \frac{8}{f^2}\left(L_5^Q-L_5\right) m_K^2\;,\label{eq:q6}
\end{split}
\end{equation}
which does not depend on counterterms if $L_5^{full}=L_5^{quench}$. 
In eq.~\eqref{eq:q6} we have neglected counterterms suppressed by $m_\pi^2/m_K^2$ 
and assumed that $\delta f$ is small ($\delta f$ is independent of counterterms 
by construction).

\section{Chiral extrapolation in $B\to \pi$ decay\label{sec:numerics2}}

\subsection{Quenched case}
So far the quenched lattice studies of the $B\to \pi$ transition matrix element 
were  confined to the region of not very light pseudoscalar meson masses 
which (from the point of view of QChPT) is almost fortunate since one stays away 
from the region in which the spurious quenched chiral logs dominate over the 
other (physical) contributions. Supposing that one manages to push the lattice 
studies toward ever lighter ``pions", the quenched chiral log will become a more 
important effect. From the expressions 
presented in Sec.~\ref{sec:chi-results} and in appendix~B, for the very light 
meson~$m_P < (v\negcdot p)$, we obtain 
\bea \label{limit-quench}
&&f_p(v\negcdot p) = {\alpha g  \over  f (v\negcdot p + \Delta^\ast)} 
    \left[ 1 + (c_\ell^Q + c_\ell^p m_P^2) \ln\big( \frac{m_P^2}{\mu^2}\big)  + 
    c_0^p + c_1^p m_P + c_2^p m_P^2 + \dots \right] \;, \nonumber\\
&&f_v(v\negcdot p) = {\alpha \over f } 
    \left[ 1 + (c_\ell^Q + c_\ell^v m_P^2) \ln\big( \frac{m_P^2}{\mu^2}\big) + 
    c_0^v + c_2^v m_P^2 + \dots \right] \;. 
\eea
The quenched chiral log term is proportional to  
\bea \label{Qlog0}
 c_\ell^Q = -{m_0^2(1+3 g^2)\over 6 (4 \pi f)^2}\;,
\eea
and it is a quenched artifact that diverges in the chiral limit. The presence of this term has an important consequence, that by approaching ever lighter $m_P$, 
the form factors 
$f_p^{quench}$ and $f_v^{quench}$ {\it increase}. This is in contrast to the full (dynamical) 
theory in which the effect is the opposite, 
i.e., the chiral logs lower the form factors for small masses, as
we shall see in the next subsection (see fig.~\ref{plot5}).      
The form factor $f_p(v\negcdot p)$ also picks a term linear in $m_P$ in this limit, which is 
yet another artifact of the quenched approximation. The accompanying coefficient 
reads 
\bea \label{Qlog1}
c_1^p=-\frac{g^2 m_0^2}{(4 \pi f)^2} \frac{ 4 \pi }{3 v\negcdot p}\;.
\eea

\subsection{Unquenched case: An illustration}

The unquenched equivalents to the expressions~(\ref{limit-quench}) 
are not straightforward to derive. The main obstacle comes from the fact that, 
instead of one mass in the integrals $I_2(M, v\negcdot p)$ and $J_1(M, v\negcdot p)$, 
one now has three masses: $M\in \{ m_\pi$, $m_K$, $m_\eta\}$. The behavior of those 
integrals depends on the sign of $1-(v\negcdot p)/M $ (see  appendix~A). The variation 
of the pion mass entails a change of $m_K$ and $m_\eta$, which  straddle  
$(v\negcdot p)$. This then changes the behavior of the integrals $I_2$ and $J_1$. 
For the $B\to \pi$ transition, we have
\bea
f_p(v\negcdot p) &=&
{\alpha g \over f \left( v\negcdot p + \Delta^\ast \right)} 
\left\{ 1 +
{1 \over (4 \pi f)^2} \left[   \nonumber g^2   \biggl(
 4 J_1(m_\pi, v\negcdot p) + 3 J_1(m_K, v\negcdot p)+ 
{2 \over 3} J_1(m_\eta, v\negcdot p)\biggr)\right.\right. \\
&& \hspace*{10mm}
 \left.  - 
{1 + 3 g^2 \over 12} \biggl( 9 m_\pi^2 \log(m_\pi^2) +
6  m_K^2 \log(m_K^2)+    m_\eta^2 \log(m_\eta^2)\biggr) \right] 
 + d_0^p + \dots \biggr\} \;,\nonumber \\
&& \\
f_v(v\negcdot p) &=& {\alpha  \over f } \left\{ 1 + {1 \over (4 \pi f)^2} \left[
{15-27 g^2 \over 12} m_\pi^2 \log(m_\pi^2) + {1-3 g^2 \over 2}  m_K^2 \log(m_K^2)  \right. \right. \nonumber \\
&& \hspace*{15mm} \biggl. \biggr. - 
{1+3g^2 \over 12} m_\eta^2 \log(m_\eta^2) + 2 I_2(m_\pi,v\negcdot p) + 
I_2(m_K,v\negcdot p) \biggr] + d_0^v + \dots \biggr\}\;,\nonumber
\eea
where $d_0^{v,p}$ is a constant and the ellipses stand for the higher order terms 
in the $m_{\pi,K,\eta}^2$ expansion. 
\begin{table}[h]
\centering 
\begin{tabular}{|c|cccc|c|}  \hline  
{\phantom{\huge{l}}}\raisebox{-.2cm}{\phantom{\Huge{j}}}
$(v\negcdot p)$ & $a^{(0)}_{v} (\gev^{1/2})$ &  $a^{(1)}_{v} (\gev^{-3/2})$  & 
$a^{(0)}_{p} (\gev^{-1/2})$ &  $a^{(1)}_{p} (\gev^{-5/2})$ & Reference \\ 
\hline 
{\phantom{\huge{l}}}\raisebox{-.2cm}{\phantom{\Huge{j}}}
$0.55$~GeV  & $2.5(2)^{+0.0}_{-0.6}$  & $1.1(2)^{+0.0}_{-0.2}$  & 
$0.9(1)^{+0.0}_{-0.2}$  & $0.7(1)^{+0.0}_{-0.1}$ & \cite{ape} \\ 
{\phantom{\huge{l}}}\raisebox{-.2cm}{\phantom{\Huge{j}}}
$0.19$~GeV  & $0.8(3)$  & $0.3(3)$   & $4.8(4)$  & $3.6(4.5)$ & \cite{jlqcd} \\ 
\hline
\end{tabular}
{\caption{\small \label{tab2} The parameters describing the linear chiral extrapolation 
for the $B\to \pi$ form factors $f^{latt}_{v,p}(v\negcdot p)$ at fixed $ (v\negcdot p)$, as indicated in
 eq.~(\ref{extrap}). The values are those obtained in the quenched lattice studies 
in refs.~\cite{ape,jlqcd}. }}
\end{table}

To exemplify the impact of the chiral logs, we will now use the existing 
(quenched) lattice results for the $B\to \pi$ form factors presented in refs.~\cite{ape} and \cite{jlqcd}, 
in which the chiral extrapolation has been made linearly, i.e.,  
\bea \label{extrap}
f^{latt}_{v,p}(v\negcdot p) = a^{(0)}_{v,p}(v\negcdot p) + a^{(1)}_{v,p}(v\negcdot p) m_P^2 \;,
\eea
The parameters $a^{(0,1)}_{v,p}$ of ref.~\cite{jlqcd} are obtained 
by fitting in the region of light pseudoscalar mesons that corresponds to 
$m_P \in (450, 800)$~MeV. Those of ref.~\cite{ape} are obtained from a fit in $m_P \in (540, 840)$~MeV. 
The numerical values are given in table~\ref{tab2}.~\footnote{We are particularly indebted to 
Tetsuya Onogi for communicating these results to us.}

\begin{figure}[t!!] 
\begin{center}
\hspace*{-8mm}\epsfig{file=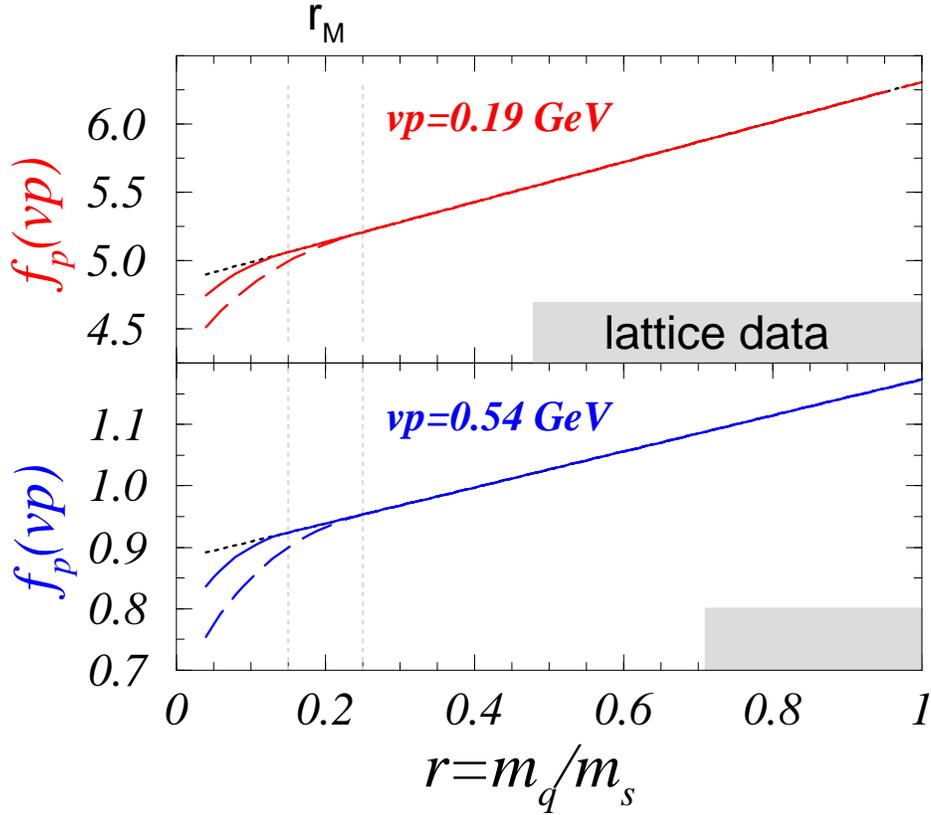,width=.77\linewidth}
\caption{\label{plot5}\footnotesize{Illustration of the effect of inclusion of 
the chiral log in extrapolating to the physical pion 
from the masses accessed from the lattice simulations (marked by the shaded boxes). 
The full (dashed) curves are obtained by including the chiral logs in extrapolation 
starting from $r_M=0.15$ and from $0.25$ (gray vertical lines) using full (unquenched) ChPT. 
For comparison we also show the (dotted) line corresponding to the linear extrapolation. }} 
\end{center}
\end{figure}

We will now assume that \underline{(a)} these results are the same in the full (unquenched) 
theory;~\footnote{Since the purpose of the discussion in this section is to illustrate 
the impact of the chiral logs on the result of an extrapolation to the physical pion mass, 
this assumption should not worry the reader.} \underline{(b)} the form~(\ref{extrap}) 
holds true down to a point $m_{M} \approx 250$~MeV or $m_{M} \approx 330$~MeV,  where we smoothly match 
the lattice and  ChPT results. The full ChPT form factor, given in appendix B, is used from the matching point $m_{M}$ 
down to the physical pion mass. In other words, for a fixed value of $(v\negcdot p)$, we take
\bea\label{hybr}
f_{p,v}(m_P^2) &=& \theta(m_P^2-m_M^2) f^{latt}_{p,v}(m_P^2)  \nonumber \\
&& +\theta(m_M^2-m_P^2) \biggl[ f^{\text{ChPT}}_{p,v}(m_P^2) - 
\left(f^{\text{ChPT}}_{p,v}(m_M^2) - f^{latt}_{p,v}(m_M^2) \right)  \biggr. \nonumber \\
&&\hspace*{22mm}\left. - \left( \left.{\partial f^{\text{ChPT}}_{p,v}/\partial m_P^2} 
\right|_{m_P^2=m_M^2} -
a^{(1)}_{v,p}\right) ( m^2_P - m^2_M) \right]\,.
\eea
In fig.~\ref{plot5} we show the effect for the form factor $f_p(v\negcdot p)$.
We observe that the form factor obtained by including the chiral logs in the extrapolation to $m_P^2=m_\pi^2$ 
is smaller than the one obtained by extrapolating linearly (dotted lines in the plot). 
The amount of that suppression obviously depends on the value of $m_M^2=4 \mu_0 m_s r_M$: the effect of the 
chiral log is smaller for smaller $r_M$. 
In our example we took  $r_M=0.15$ ($m_M \simeq  250$~MeV) and  $r_M=0.25$ ($m_M \simeq  330$~MeV).
Using  
\bea
\err(f_{p,v}) = {f_{p,v}^{eq.(\ref{hybr})} - f_{p,v}^{eq.(\ref{extrap})}\over f_{p,v}^{eq.(\ref{extrap})}} ,
\eea 
to measure the error due to chiral logs that were not included in the extrapolation to the physical pion mass, we
obtain the following results.
\bea
\underline{\sf r_M=0.15} \hspace*{6.5cm} && \cr
\err(f_{p}(v\negcdot p=0.19\; \gev)) \simeq -2\% \;,\quad  & & 
\err(f_{p}(v\negcdot p=0.54\; \gev)) \simeq -5\% \;,\cr
&&\cr
 \underline{\sf r_M=0.25} \hspace*{6.5cm}  && \cr
\err(f_{p}(v\negcdot p=0.19\; \gev)) \simeq -7\% \;,\quad  & & 
\err(f_{p}(v\negcdot p=0.54\; \gev)) \simeq -15\% \;.
\eea
The above analysis applied to the form factor $f_v$
leads to even smaller uncertainties: 
$\err(f_v)_{r_M=0.15} < 3 \%
$ and 
$\err(f_v)_{r_M=0.25} < 6 \%
$, for both values of $(v\negcdot p)$.

This exercise is made just to illustrate how one can proceed in order to get an 
estimate of the syste\-ma\-tic uncertainties due to the chiral extrapolation. As
we saw, the amount of estimated uncertainty is highly sensitive to the 
choice of the point $r_M$, which is why the outcome of this exercise 
should be considered only as a rough estimate.

It is important to stress again that had we used the quenched 
expressions~\eqref{limit-quench} to guide the chiral extrapolation, the 
result would stay above the result of the linear extrapolation, precisely the 
opposite to what happens in the full theory (which we show in fig.~\ref{plot5}).

\begin{figure}
\vspace*{-5mm}
\begin{center}
\hspace*{-8mm}\epsfig{file=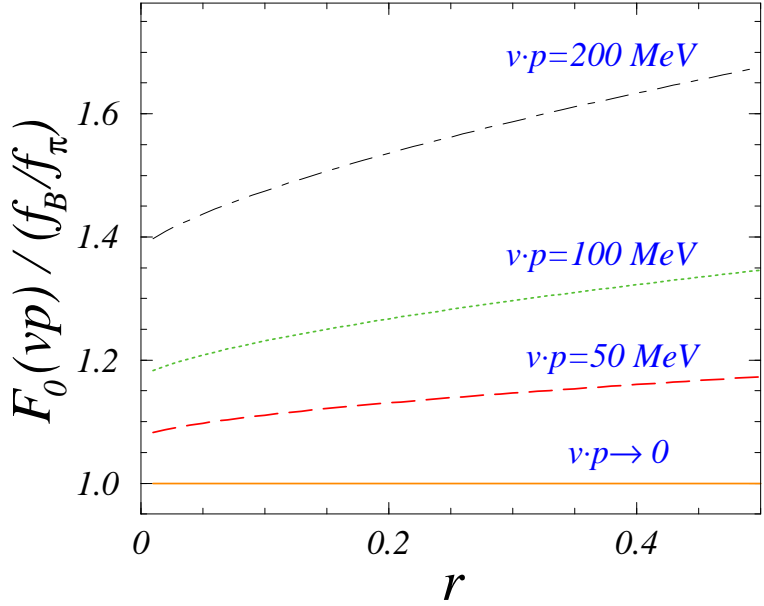,width=.62\linewidth}
\vspace*{-5mm}
\caption{\label{fig:SPT}\footnotesize{The ratio $F_0(v\negcdot p)/(f_B/f_\pi)$ 
or $f_v/(\hat f/f)$, which satisfies the soft pion theorem at zero recoil 
($v\negcdot p \to 0$). The figure shows the ratio in the dynamical theory at which 
the violation of the soft pion theorem 
grows fast with the recoil momentum. The illustration is provided for three momenta and 
for masses corresponding to $r\in (0,0.5)$, i.e., $m_\pi \in (0,0.5)$~GeV. Notice that 
in the quenched theory this ratio does not depend on $(v\negcdot p)$ and is equal to $1$.}} 
\end{center}
\end{figure}

Finally we would like to make a comment on the ratio  $f_v/(\hat f/f)$ or 
equivalently $F_0(v\negcdot p)/(f_B/f_\pi)$, which, according to the soft pion theorem, 
should be equal to $1$ at the zero recoil point $v\negcdot p \to 0$. In the 
quenched theory the form factor $F_0$ is independent of 
$v\negcdot p$ and its chiral corrections are exactly the same as in $f_B/f_\pi$. 
Therefore the ratio is indeed $1$ for all combinations of $(r, v\negcdot p)$. 
In the full (unquenched) theory, in contrast, the chiral corrections cancel only at 
$v\negcdot p \to 0$ and the soft pion theorem is satisfied. It is worth noticing,
however, that when a small recoil is introduced the violation of the soft pion 
theorem is quite large in the full theory (see fig.~\ref{fig:SPT}).

\section{Conclusions\label{sec:conclusions}}

In this paper we explored an approach that contains at its core the leading order in the heavy 
quark expansion and the next-to-leading order in ChPT, to derive expressions for the heavy-to-light 
pseudoscalar 
semileptonic form factors. The expressions are worked out in both standard and quenched ChPT. 
In the latter case we observe a familiar feature of  quenched calculations, namely, the 
appearance of  divergent chiral logarithms (quenched chiral logs), which makes it harder to 
compare the results with the formulas obtained in standard ChPT for $B\to \pi$ form factors: 
one should find a fiducial window in  which the masses are not so light so that the quenched 
chiral logs do not dominate the QChPT expressions, yet small enough for the chiral 
expansion to be meaningful. Furthermore, for the numerical estimates a number 
of low energy constants must be fixed by using both the available experimental data and the results 
of quenched lattice simulations. Our numerical estimates in which we use the QChPT expressions 
are highly sensitive to the value of the parameter $g^\prime$ (the coupling of the doublet of 
heavy-light mesons to a light $\eta^\prime$ meson). Until a lattice computation of that 
parameter is made, we are not able to make firm quantitative statements.  
Even information about the sign of $g^\prime$ would be welcome.   
It turns out that, for $g^\prime = - g$ (which, on the basis of the large $N_c$ expansion, 
is the limiting value), we get a less pessimistic scenario: 
one can even find  combinations of $( v\negcdot p)$ and the light meson mass $m_P$ 
such that the quenching errors on $F_{+,T}(q^2)$ vanish. 
We were also able to find the combinations of 
$( v\negcdot p)$ and $m_P$ such that the quenching errors do not depend on the value of 
$g^\prime$; at these points and for small recoils the quenching errors are between 
$15\%
$ and $25\%
$. We reiterate that the numerical estimates do depend  on the specific choice of 
the low energy constants.

As for the form factor $F_0(q^2)$, the present approach suggests that 
the quenching errors are large regardless of the value of the 
coupling $g^\prime$, the quenched 
value being larger than the unquenched one. Only at the point
corresponding to zero recoil are those errors reasonably small. Away 
from that point, they are large.

The same observations apply also when the final meson is a kaon. In that 
case, by using the QChPT expressions, we were able to verify that the form factors 
obtained for the kaon consisting of degenerate and nondegenerate quarks are 
effectively indistinguishable, provided the value of the coupling is $g^\prime = -g$. 
For  $g^\prime=+g$ these uncertainties also become large.

Finally, the formulas presented in this work may be useful in assessing the 
systematic uncertainty due to chiral extrapolations. 
We showed how to include the chiral logs to extrapolate from the region in 
which the pion is heavier than $400$~MeV. As could have been anticipated, 
the estimated uncertainty of the chiral extrapolation  depends on the 
mass from which the chiral logs are included in the extrapolation.

We also provided the quenched chiral log coefficient, which might be useful 
if lattice calculations are performed with very light mesons.

We verified that the ratio $F_0^{B\to \pi}/(f_B/f_\pi)$ satisfies the soft pion theorem, 
i.e., it is equal to $1$ at zero recoil, in both theories. In the quenched theory that value 
remains unchanged even when a small recoil momentum is introduced. In the dynamical theory, 
instead, this ratio is significantly larger than $1$ away from zero recoil.

\section*{Acknowledgments}
We thank T.~Onogi and G.~Lacagnina for communicating the details of their work to us,   
S.~Fajfer, J.~Flynn, V.~Lubicz, G.~Martinelli, and S.~Sharpe for their comments on the manuscript, 
and Suzy~Vascotto for her help in polishing the text. 
The work of S.P. and J.Z. was supported in part by the Ministry 
of Education, Science and Sport of the Republic of Slovenia.

\appendix
\newpage 
\refstepcounter{section}
\section*{Appendix A: Chiral loop integrals}\label{appA}

In this appendix we list the dimensionally regularized integrals encountered 
in the course of the calculation. For more details, see~\cite{jure} and references therein: 
\begin{align} \label{int-A}
i\mu^\epsilon\int \frac{d^{4-\epsilon}q}{(2\pi)^{4-\epsilon}
}\frac{1}{q^2-m^2}&=\frac{1}{16 \pi^2}I_1(m)\;, \nonumber \\
i\mu^\epsilon\int \frac{d^{4-\epsilon}q}{(2\pi)^{4-\epsilon}}
\frac{1}{(q^2-m^2)(q\negcdot v-\Delta)}&=\frac{1}{16
\pi^2}\frac{1}{\Delta}I_2(m,\Delta)\;,
\end{align}
where
\begin{align}
I_1(m)&=m^2 \ln\Bigl(\frac{m^2}{\mu^2}\Bigr)-m^2\bar{\Delta}\; , \nonumber \\
I_2(m,\Delta)&=-2\Delta^2 \ln\Bigl(\frac{m^2}{\mu^2}\Bigr)-4 \Delta^2
F\Bigl(\frac{m}{\Delta}\Bigr) +2 \Delta^2(1+\bar{\Delta})\; ,
\end{align}
where $\bar{\Delta}=2/\epsilon -\gamma +\ln(4\pi)+1$. The function $F(x)$ was 
 calculated in ref.~\cite{stewart}, for both  negative and positive values of
the argument:
\begin{equation}
F\left(\frac{1}{x}\right)= \left\{
\begin{aligned}
-\frac{\sqrt{1-x^2}}{x}&\left[\frac{\pi}{2}-\tan^{-1}\left(\frac{x}{\sqrt{1-x^2}}\right)\right]\,,\qquad
&|x|\le 1\,,\\
\frac{\sqrt{x^2-1}}{x}&\ln \left(x+\sqrt{x^2-1}\right)\,,\qquad &|x|\ge 1
\; .
\end{aligned}\right.
\end{equation}
In addition to the integrals~(\ref{int-A}), one also needs the following two:
\bea
&&i\mu^\epsilon\int \frac{d^{4-\epsilon}q}{(2\pi)^{4-\epsilon}}
\frac{q^\mu}{(q^2-m^2)(q\negcdot v-\Delta)}=\frac{v^\mu}{16
\pi^2} \left[ I_2(m,\Delta)+I_1(m) \right]\;, \nonumber \\
&& i\mu^\epsilon\int \frac{d^{4-\epsilon}q}{(2\pi)^{4-\epsilon}}
\frac{q^\mu q^\nu}{(q^2-m^2)(q\negcdot v-\Delta)}=\frac{1}{16
\pi^2}\Delta\left[ J_1(m,\Delta)\eta^{\mu \nu}+J_2(m,\Delta)v^\mu
v^\nu \right] \;,
\eea
with
\begin{subequations}\label{eq-1}
\begin{align}
\begin{split}
J_1(m,\Delta)=(-m^2&+\frac{2}{3}\Delta^2)\ln\left(\frac{m^2}{\mu^2}\right)+\frac{4}{3}(\Delta^2-m^2)
F\left(\frac{m}{\Delta}\right)\\
&\qquad\qquad -
\frac{2}{3}\Delta^2(1+\bar{\Delta})+\frac{1}{3}m^2(2+3\bar{\Delta})+\frac{2}{3}m^2-\frac{4}{9}\Delta^2
\; ,
\end{split}
\\
\begin{split}
J_2(m,\Delta)=(2
m^2&-\frac{8}{3}\Delta^2)\ln\left(\frac{m^2}{\mu^2}\right)-\frac{4}{3}(4
\Delta^2-m^2)F\left(\frac{m}{\Delta}\right)\\
&\qquad\qquad+\frac{8}{3}\Delta^2(1+\bar{\Delta})-\frac{2}{3}m^2(1+3\bar{\Delta})-\frac{2}{3}m^2+
\frac{4}{9}\Delta^2
\; .
\end{split}
\end{align}
\end{subequations}
The functions $J_1(m,\Delta),J_2(m,\Delta)$ differ from the ones in
ref.~\cite{boyd} by the last two terms in
\eqref{eq-1} which are of ${\cal O}(m^2, \Delta^2)$. These
additional (finite) terms originate from the fact that $\eta^{\mu \nu}$ is
the ($4-\epsilon$)-dimensional metric tensor.

\section*{Appendix B: Explicit expressions for the one-loop chiral corrections}\refstepcounter{section}\label{appB}

As already mentioned in the body of the paper, the chiral loop corrections to the 
form factors can be written in the form 
\bea
\delta f_{p,v} = \sum_{I} \delta f_{p,v}^{(I)} + \frac{1}{2} \delta Z_B^{\text{loop}} + \frac{1}{2} \delta Z_P^{\text{loop}}\,,
\eea
where the sum goes over all the graphs depicted in fig.~\ref{fig2}.  In what follows
we give the explicit expressions for both form factors graph by graph and in both chiral 
theories (quenched and standard).

\subsection{Quenched theory}\label{appBQuenched}

In the calculation of one-loop contributions, we made several 
approximations in order to simplify the final expressions. We make 
use of the fact that $v\negcdot p > \Delta^{\ast}$ for the $B\to P$ 
transition and thus  consistently neglect 
the mass differences between $B$, $B^\ast$, $B_s$, and $B_s^\ast$ mesons in the loops. 
This neglect induces a spurious singularity in the expression for the diagrams 
(7a) and (7b) in fig.~\ref{fig2}, at $v\negcdot p \to 0$. To handle those singularities we follow the 
proposal by Falk and Grinstein~\cite{falk} and resum the corresponding diagrams 
and then subtract the term that would renormalize the $B^\ast$ meson mass. 
We recall that $a$ $(b)$ superscripts distinguish the diagrams without (with) hairpin 
insertion. The  formulas for the $B_j\to P_{ij}$ transition (with the valence 
quark content of mesons $B_j\sim b\bar q_j$ and $P_{ij}\sim q_i\bar q_j$) 
are expressed in terms of the  pseudoscalar meson mass $M_i^2 = 4 \mu_0 m_i$: 
\bea 
&&\delta f_p^{(7a)} = {6 g g'\over (4 \pi f)^2}\    
\left[  J_1(M_i, v\negcdot p)-\frac{1}{v\negcdot p} \frac{2\pi}{3} M_i^3\right] \;, \nonumber \\
&&\delta f_p^{(7b)} = - { g^2 \over (4 \pi f)^2}
 \left[ \alpha_0 + \Big(\alpha_0
M_i^2-m_0^2\Big)\frac{\partial}{\partial M_i^2}\right] 
\Big( J_1(M_i, v\negcdot p)-\frac{1}{v\negcdot p} \frac{2\pi}{3} M_i^3\Big) \;,\nonumber \\
&&\delta f_p^{(9a)}= { g g^\prime \over (4 \pi f)^2}
\bigg[  J_1(M_i,v\negcdot p)+J_1(M_j,v\negcdot p) -\frac{1}{v\negcdot p}\frac{2 \pi}{3}\Big(M_i^3+M_j^3\Big)
\bigg]\;,\nonumber \\
&& \delta f_p^{(9b)} = {g^2 \over 3 (4 \pi f)^2 } 
 \biggl\{ {\alpha_0 M_j^2-m_0^2\over M_j^2-M_i^2}
\Big[\frac{1}{v\negcdot p}\frac{2\pi}{3} M_j^3- J_1(M_j,v\negcdot p)\Big] \biggr.\nonumber \\
&&  \hspace*{37mm}\biggl.- { \alpha_0 M_i^2-m_0^2 \over M_j^2-M_i^2 } 
\Big[\frac{1}{v\negcdot p}\frac{2\pi}{3}M_i^3- J_1(M_i,v\negcdot p)\Big]\biggr\}\;, \nonumber \\
&& \delta f_p^{(12b)} =\frac{T}{18 (4 \pi f)^2} \biggl\{ 
\frac{ 2 m_0^2-\alpha_0 (M_i^2 + M_j^2)}{ M_j^2-M_i^2 }
\Big[I_1(M_j)-I_1(M_i)\Big] \biggr. \nonumber \\ \biggl. 
&& \hspace*{37mm}+\Big(\alpha_0 M_i^2-m_0^2\Big)
\frac{\partial I_1(M_i)}{\partial M_i^2}+\Big(\alpha_0 M_j^2-m_0^2\Big)
\frac{\partial I_1(M_j)}{\partial M_j^2}\biggr\} \;, \nonumber \\
&& \delta f_p^{(13a)} =- \frac{i V_L'(0)f}{\sqrt{ 6} (4 \pi f)^2} I_1(M_i)\;, \nonumber \eea \bea
&&\delta f_p^{(13b)} = \frac{1}{6 (4 \pi f)^2}
\Big[\alpha_0 I_1(M_i)+\Big(\alpha_0 M_i^2-m_0^2\Big)\frac{\partial I_1(M_i)}
{\partial M_i^2} \Big]\,,\label{fpQ}
\eea
where $T=0$ for $i=j$, and $T=1$ otherwise.
The functions $I_1(m)$, $J_1(m,\Delta)$ are given in appendix~\ref{appA}. 
As for the form factor $f_v(v\negcdot p)$, the nonzero chiral loop corrections are
\begin{align}
\delta f_v^{(4a)}&=-\frac{i T V_L^\prime(0) f}{\sqrt{6} (4 \pi f)^2} \biggl[ 
 I_2(M_j,v\negcdot p)-I_2(M_i,v\negcdot p) + {1\over 2} \left[I_1(M_j)-I_1(M_i)\right] \biggr]\,, \nonumber
\\
\begin{split}
\delta f_v^{(4b)}&=\frac{T}{ 6 (4 \pi f)^2} \biggl\{\frac{1}{M_j^2-M_i^2}\Big[\left(\alpha_0
M_j^2-m_0^2\right)[I_1(M_j)+2 I_2(M_j,v \negcdot p )]\\
&\hspace*{37mm}-(\alpha_0 M_i^2-m_0^2) \left( I_1(M_i)+
2 I_2(M_i,v \negcdot p ) \right) \Big]\\
&\qquad\qquad - \left(\alpha_0 +(\alpha_0 M_i^2-m_0^2)\frac{\partial} {\partial M_i^2}\right)  
\left(I_1(M_i)+2I_2(M_i,v\negcdot p)\right) \biggr\}\,,\\
\delta f_v^{(14a)}&=-\frac{i V_L'(0)f}{2\sqrt{6} (4 \pi f)^2 } 
\left( I_1(M_i)+I_1(M_j)\right) \,,
\end{split}
 \nonumber \\
\begin{split}
\delta f_v^{(14b)}&=\frac{1}{18 (4 \pi f)^2}\biggl\{ 
(\alpha_0 M_j^2-m_0^2)\Big[\frac{I_1(M_j)}{M_j^2-M_i^2} + 
\frac{\partial I_1(M_j)}{\partial M_j^2} \Big] \\
&\hspace*{19mm} + (\alpha_0 M_i^2-m_0^2)\Big[\frac{I_1(M_i)}{M_i^2-M_j^2} + 
\frac{\partial I_1(M_i)}{\partial M_i^2} \Big] + \alpha_0 \left[ 
I_1(M_i) + I_1(M_j) \right] \biggr\}\,.
\end{split}\label{fvQ}
\end{align}
In the wave function renormalization factors $Z_{B,P}$, we separate
the one-loop chiral contribution $\delta Z^{\text{loop}}_{B,P}$ from the pieces  
coming from the counterterms $\delta Z^{c.t.}_{B,P}$, i.e.,
\begin{equation}
Z_{B,P}=1+\delta Z_{B,P}=1+ \delta Z^{\text{loop}}_{B,P}+\delta Z^{c.t.}_{B,P}\label{deltaZ}\;.
\end{equation}
More specifically,
\begin{subequations}
\begin{align}
\delta Z_{B_j}^{\text{loop}}&= {1\over (4 \pi f)^2} \Big[(2 g^2 \alpha_0 M_j^2-6 g g'M_j^2-g^2 m_0^2) \ln\Big(\frac{M_j^2}{\mu^2}
\Big)\ ,\nonumber\\
&+\alpha_0 g^2M_j^2-m_0^2 g^2 +\Big(-2g^2M_j^2\alpha_0+6 g g' M_j^2 +g^2
m_0^2 \Big)\bar{\Delta}\Big]\\
\delta Z_{P_{ij}}^{\text{loop}}&= {1\over 9 (4 \pi f)^2} \Bigl\{  
\frac{\ln\big(M_j^2/M_i^2\big)}{M_j^2-M_i^2}\big[ 2 \alpha_0 M_j^2 M_i^2- m_0^2 (M_j^2+M_i^2)\big]\nonumber\\
&\qquad + 2 m_0^2  -\alpha_0 (M_i^2+M_j^2)  \Bigr\} ~,
\end{align}
\end{subequations}
while the counterterms contribute as 
\begin{equation}
\delta Z_{B_j}^{c.t.}= k_1 m_j\;, \hskip2cm 
\delta Z_{P_{ij}}^{c.t.}=-8 L_5\frac{4\mu_0}{f^2}(  m_{i} +m_{j})\;.
\end{equation}
Thus the wave function renormalization factor for $\pi$ and $K$ mesons reads 
\begin{subequations}
\begin{align}
Z_\pi&=1 -8 L_5\frac{4\mu_0}{f^2}~2  m_q\,,\\
Z_K&=1 -\frac{1}{9 (4 \pi f)^2 } \bigg\{ \frac{\ln \big({ m_s/m_q } \big)}
{(m_K^2-m_\pi^2)} \Big(\alpha_0 m_\pi^4-\alpha_0 2 m_K^2 m_\pi^2 +m_0^2 m_K^2\Big)\nonumber\\
&\qquad \quad  + 2 \alpha_0 m_K^2 - 2 m_0^2  \Big\} -8 L_5\frac{4 \mu_0}{f^2}( m_q +m_s)~,
\end{align}
where we ignore  the isospin-breaking effects and set $m_u=m_d=m_q$.
\end{subequations}

Finally, we also display the expression for the heavy-light meson decay 
constant: 
\bea
&&f_{B_i}=\frac{\alpha}{\sqrt{m_B}}\Biggl\{
1+\frac{1}{6 (4 \pi f)^2}
\Bigl[ \left[\alpha_0 - if\sqrt{6}V_L^\prime(0) \right] I_1(M_i)+
 (\alpha_0 M_i^2-m_0^2)\frac{\partial I_1(M_i)}{\partial M_i^2}\Bigr]\nonumber \\
&&\hspace*{28mm}+\varkappa_1m_i+\tfrac{1}{2}\delta Z_{B_i}\Biggr\}\,,
\eea
in agreement with refs.~\cite{booth,zhang}.

\subsection{Full (unquenched) theory}\label{appBUnQuenched}

In this subsection we present the expressions for the form factors in  
the full theory. The nonanalytic contributions to the form factors 
in this theory have already been calculated in ref.~\cite{falk}. Our 
results also include the analytic terms. As in the quenched case, we 
work in  the isospin limit $m_u=m_d=m_q$ and neglect  the 
differences of heavy meson masses in the loops. We now list the results for 
$\delta f_{v,p}^{(I)}$ in  $B_j \to P_{ij}$ mediated by the $(V-A)$ operator, where
$P_{ij}$ stands for the light pseudoscalar meson with the valence quark content 
$\bar q_i q_j$.

The form factor $f_{p}(v\negcdot p)$ receives the following one-loop corrections:
\begin{align}
\delta f_p^{(7a)}&=\frac{3 g^2}{(4 \pi f)^2}\Biggl\{ \sum_{P'} C_{B_j P' P_{ij}}^{(7a)}  \left[ 
J_1(m_{P'}, v\negcdot p )-\frac{1}{v\negcdot p}\frac{2 \pi}{3} m_{P'}^3 \right] \Biggr\} \;,\nonumber \\
\delta f_p^{(9a)}&=-\frac{g^2}{(4 \pi f)^2}\Big\{ \sum_{P'} C_{B_j P' P_{ij}}^{(9a)}\Big[ 
  J_1(m_{P'}, v\negcdot p ) - \frac{1}{v\negcdot p}\frac{2 \pi}{3} m_{P'}^3\Big]\Big\} \;,\nonumber \\
\delta f_p^{(12a)}&= \frac{1}{(4 \pi f)^2} \left[\sum_{P'} C_{B_j P' P_{ij}}^{(12a)}I_1(m_{P'})\right] \;,\nonumber 
\\
\delta f_p^{(13a)}&= \frac{1}{(4 \pi f)^2} \left[\sum_{P'} C_{B_j P'
P_{ij}}^{(13a)}I_1(m_{P'})\right] \;,
\end{align} 
where the coefficients $C_{B_j P' P_{ij}}$ depend on the final and initial states. 
A detailed list of coefficients includes the following.
\begin{itemize}
\item[$\circ$] {\sc for the $B \to K $ transition,} \\
\begin{tabular}{l l l l l l l}
$C_{B \pi K}^{(7a)}=0$, & $C_{B K  K}^{(7a)}=2$, &$C_{B \eta K}^{(7a)}=2/3$;&\qquad&$
C_{B \pi K}^{(9a)}=0$, &$C_{B K  K}^{(9a)}=0$, &$C_{B \eta K}^{(9a)}=\frac{1}{3}$;
\\ 
$C_{B \pi K}^{(12a)}=-\frac{1}{4}$, &$C_{B K K}^{(12a)}=-\frac{1}{2}$, &$C_{B \eta K}^{(12a)}=-\frac{1}{4}$;&\qquad&
$C_{B \pi K}^{(13a)}=0$, &$C_{B K K}^{(13a)}=-1$, &$C_{B \eta K}^{(13a)}=-\frac{1}{3}$.
\end{tabular}

\item[$\circ$] {\sc   for the $B\to \pi$ transition,} \\
\begin{tabular}{l l l l l l l}
 $C_{B \pi \pi}^{(7a)}=\frac{3}{2}$,& $C_{B K  \pi}^{(7a)}=1$,& $C_{B \eta \pi}^{(7a)}=\frac{1}{6}$;&\qquad&
 $C_{B \pi \pi}^{(9a)}=\frac{1}{2}$,& $C_{B K  \pi}^{(9a)}=0$,& $C_{B \eta \pi}^{(9a)}=-\frac{1}{6}$;
\\ 
$C_{B \pi \pi}^{(12a)}=-\frac{2}{3}$,& $C_{B K \pi}^{(12a)}=-\frac{1}{3}$,&$C_{B \eta \pi}^{(12a)}=0$;&\qquad&
$C_{B \pi \pi}^{(13a)}=-\frac{3}{4}$,& $C_{B K \pi}^{(13a)}=-\frac{1}{2}$,&
$C_{B \eta \pi}^{(13a)}=-\frac{1}{12}$. 
\end{tabular}

\item[$\circ$] {\sc   for the $B_s \to K$ transition,} \\
\begin{tabular}{l l l l l l }
$C_{B_s \pi K}^{(7a)}=\frac{3}{2}$,& $C_{B_s K  K}^{(7a)}=1$,&$C_{B_s \eta K}^{(7a)}=\frac{1}{6}$; &
$C_{B_s \pi K}^{(9a)}=0$,& $C_{B_s K  K}^{(9a)}=0$,& $C_{B_s \eta K}^{(9a)}=\frac{1}{3}$;\\
 $C_{B_s \pi K}^{(12a)}=-\frac{1}{4}$,&$C_{B_s K K}^{(12a)}=-\frac{1}{2}$,& $C_{B_s \eta K}^{(12a)}=-\frac{1}{4}$;&
$C_{B_s \pi K}^{(13a)}=-\frac{3}{4}$,& $C_{B_s K K}^{(13a)}=-\frac{1}{2}$,& 
$C_{B_s \eta K}^{(13a)}=-\frac{1}{12}$.
\end{tabular}
\end{itemize}
\vskip5mm
\noindent
The  non-vanishing one-loop chiral corrections to $f_v(v\negcdot p)$ form factor are
\begin{align}
\delta f_v^{(4a)}&= \frac{1}{(4\pi f)^2}\Big\{ \sum_{P'} D_{B_jP' P_{ij}}^{(4a)}\Big[
I_2(m_{P'},v\negcdot p)+\frac{1}{2}I_1(m_{P'})\Big]\Big\} \;, \nonumber \\
\delta f_v^{(14a)}&=\frac{1}{(4\pi f)^2} \left[\sum_{P'}D_{B_jP'
P_{ij}}^{(14a)}I_1(m_{P'})\right] \;,
\end{align}
where the process-dependent coefficients $D_{B_jP' P_{ij}}$ are as follows. 
\begin{itemize}
\item [$\circ$] {\sc for  the $B\to K$ transition,}\\
$\begin{aligned}
D_{B\pi K}^{(4a)}&=0 , D_{B K K}^{(4a)}=2 , D_{B \eta  K}^{(4a)}=1;& 
D_{B\pi K}^{(14a)}=-1/4 , D_{B K K}^{(14a)}=-1/2 , D_{B \eta  K}^{(14a)}=-1/12.&
\end{aligned}$

\item[$\circ$] {\sc  for the $B\to \pi $ transition,}\\
$\begin{aligned}
 D_{B\pi \pi}^{(4a)}&=2  ,  D_{B K \pi}^{(4a)}=1  ,  D_{B \eta  \pi}^{(4a)}=0 ;&  
 D_{B\pi \pi}^{(14a)}=-5/12  ,  D_{B K \pi}^{(14a)}=-1/3  ,  
 D_{B \eta  \pi}^{(14a)}=-1/12 .&
\end{aligned}$

\item[$\circ$] {\sc  for the $B_s\to K$ transition,}\\
$D_{B_s\pi K}^{(4a)}=3/2$ , $D_{B_s K K}^{(4a)}=1$ , $D_{B_s \eta  K}^{(4a)}=1/2$; 
$D_{B_s\pi K}^{(14a)}=-1/4$ , $D_{B_s K K}^{(14a)}=-1/2$ , $D_{B_s \eta  K}^{(14a)}=-1/12$.
\end{itemize}
\vskip5mm
\noindent
The wave function renormalization factors $Z$ for $B$ mesons in the full theory are
\begin{align}
Z_{B_{u,d}}&= 1-\frac{3g^2}{(4 \pi f)^2}\left[ \frac{3}{2}
I_1(m_\pi)+I_1(m_K)+\frac{1}{6}I_1(m_\eta)\right] + k_1  m_q + k_2 (m_u+m_d+m_s)\;, \nonumber \\
Z_{B_s}&=1-\frac{3g^2}{(4 \pi f)^2} \left[ 2 I_1(m_K)+ \frac{2}{3} I_1(m_\eta)\right] + k_1
m_s + k_2 (m_u+m_d+m_s)\;,
\end{align}
while for  light mesons we have
\begin{align}
\begin{split}
Z_{K}&=1+\frac{1}{(4 \pi f)^2}
\left[\frac{1}{2}I_1(m_\pi)+I_1(m_K)+\frac{1}{2}I_1(m_\eta)\right] \\
&\qquad\qquad\qquad\qquad
-8L_5\frac{4\mu_0}{f^2}
\left( m_q +m_s\right)-16L_4\frac{4\mu_0}{f^2}\left( m_u+m_d+m_s \right)\;,
\end{split} \nonumber \\
Z_{\pi}&=1+\frac{2}{3(4 \pi f)^2} \left[  2 I_1(m_\pi) + I_1(m_K)\right] -
8 L_5\frac{4\mu_0}{f^2}~2  m_q -
16 L_4\frac{4\mu_0}{f^2} \left( m_u+m_d+m_s \right)\;.
\end{align}
As in the previous subsection we close the list of results by showing also the 
corresponding formulas for the heavy-light decay constants. We have
\begin{align}
f_{B_s}&=\frac{\alpha}{\sqrt{m_B}}\Big\{1-\frac{1}{(4 \pi f)^2}\big[I_1(m_K)+\frac{1}{3}I_1(m_\eta)\big]+
\varkappa_1m_s+\varkappa_2(m_u+m_d+m_s)+\frac{1}{2}\delta Z_{B_s}\Big\}\;, \nonumber \\
\begin{split}
f_{B_{u,d}}&=\frac{\alpha}{\sqrt{m_B}}\Big\{1-\frac{1}{(4 \pi f)^2}\big[\frac{3}{4}I_1(m_\pi)+\frac{1}
{2}I_1(m_K)+\frac{1}{12}I_1(m_\eta)\big]\\
&\qquad\qquad\qquad\qquad\qquad\qquad\qquad\qquad
+\varkappa_1  m_q +\varkappa_2(m_u+m_d+m_s)+\frac{1}{2}\delta{Z_{B_{u,d}}}\Big\}\;,
\end{split}
\end{align}
in agreement with the results of refs.~\cite{booth,zhang}.

\end{document}